\documentclass[sigconf]{acmart}

\AtBeginDocument{%
  }

\usepackage{subcaption}
\usepackage{algorithmic}
\usepackage{textcomp}
\usepackage{multirow}
\usepackage{pifont}
\usepackage{listings}
\usepackage{enumitem}

\newcommand{\llmplaceholder}[1]{\texttt{\{#1\}}}

\usepackage{framed}
\newenvironment{llmscriptblock}[1]{%
    \medskip
    \vspace{-0.5em}
    \begin{snugshade*}
    \small\ttfamily 
}{%
    \end{snugshade*}
    \medskip
}
\definecolor{promptgray}{gray}{0.95}
\colorlet{shadecolor}{promptgray}

\newcommand{\cmark}{\ding{51}}
\newcommand{\xmark}{\ding{55}}

\copyrightyear{2026}
\acmYear{2026}
\setcopyright{cc}
\setcctype{by}
\acmConference[DIS '26]{Designing Interactive Systems Conference}{June 13--17, 2026}{Singapore, Singapore}
\acmBooktitle{Designing Interactive Systems Conference (DIS '26), June 13--17, 2026, Singapore, Singapore}
\acmDOI{10.1145/3800645.3813081}
\acmISBN{979-8-4007-2563-0/2026/06}

\newcommand{\papername}{\textsc{TaskLens}}

\begin{document}

\title{\papername{}: Generating Task-Conditioned Scaffolded Interfaces for Learning Professional Creative Software}

\author{Yimeng Liu}
\email{yimengliu@ucsb.edu}
\affiliation{%
  \institution{University of California, Santa Barbara}
  \city{Santa Barbara}
  \state{California}
  \country{USA}
}

\author{Misha Sra}
\email{sra@ucsb.edu}
\affiliation{%
  \institution{University of California, Santa Barbara}
  \city{Santa Barbara}
  \state{California}
  \country{USA}
}

\renewcommand{\shortauthors}{Liu et al.}

\begin{abstract}
Professional creative software has steep learning curves for novices due to complex interfaces, limited guidance, and unfamiliar terminology. To support educators and tool creators in addressing learner challenges, we introduce \papername{}, an LLM-based method that automatically generates task-conditioned scaffolded UIs from natural language task descriptions. Our method uses LLMs to identify workflow stages and domain concepts, select task-relevant tools, generate implementation code, and execute the code to produce scaffolded interfaces. The interfaces surface relevant tools, organize them by workflow stage, link them to domain concepts, and progressively disclose advanced features. We evaluate \papername{} by deploying two LLM-generated scaffolded interfaces in Blender, a professional 3D modeling software. A user study with beginners (n=32) showed that our scaffolded interfaces significantly reduced perceived task load, improved task performance through embedded workflow guidance, and increased domain concept learning in Blender during task execution. A second study with experts (n=8) showed improved task efficiency and potential to create personalized UIs for productivity and creativity. 
\end{abstract}

\ccsdesc[500]{Human-centered computing~Interactive systems and tools}

\keywords{professional creative software, user interface generation, scaffolded learning, task assistance}

\begin{teaserfigure}
  \includegraphics[width=\textwidth]{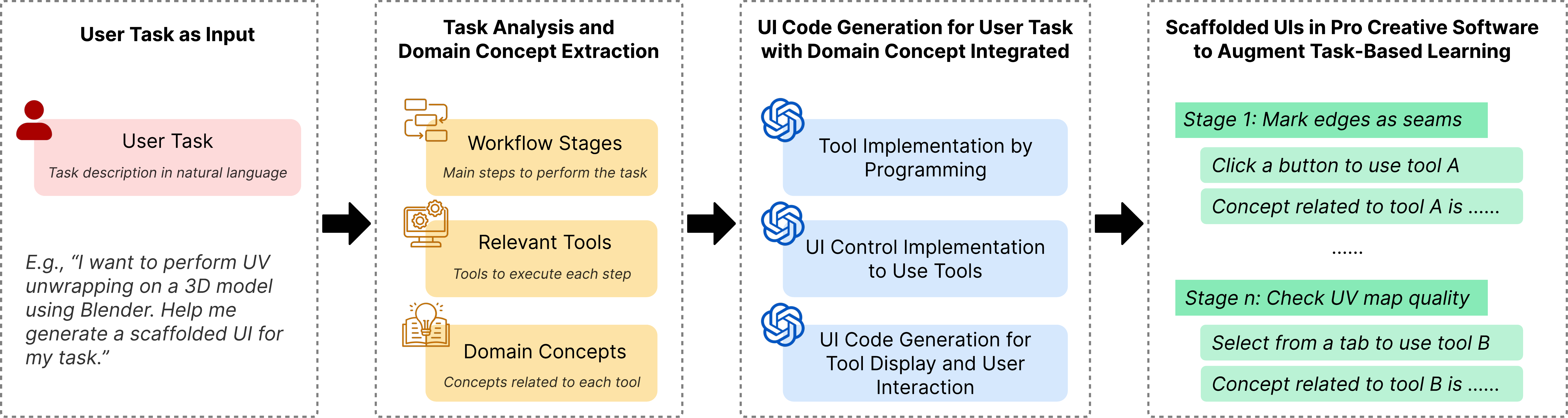}
  \caption{With \papername{}, users can generate task-conditioned scaffolded user interfaces (UIs) to augment software and domain concept learning while executing their tasks in professional creative software. Taking a task description as input, \papername{} automatically analyzes user task to decompose the task into workflow stages and extract relevant tools and domain concepts. The functionality of selected tools are implemented through programming and presented to the user as UI controls on the scaffolded interface. The code to implement functionality and UI controls are generated by LLMs and executed in the software for user interaction. Users can interact with the scaffolded interface with task-relevant tools structured around the workflow stages and learn domain concepts through labeled tooltips during task execution.}
  \Description{This figure shows a flowchart for scaffolded UI generation. Starting from the left, the flowchart takes user task as input, performs task analysis and domain concept extraction, generates UI implementation code with LLMs, and executes the code in professional creative software for user interaction.}
  \label{fig:teaser}
\end{teaserfigure}

\maketitle

\section{Introduction}
Professional creative software, such as Photoshop~\cite{photoshop}, Illustrator~\cite{illustrator}, and Blender~\cite{blender}, offers significant skill augmentation and forms the backbone of many creative and design industries. However, this skill comes at a cost of increased learning complexity. Prior work has identified that a major complexity arises from the user interface (UI), where newcomers often encounter hundreds of icons, menus, and specialized tools~\cite{chilana2018supporting}. Learners frequently struggle to find the appropriate functions or tools and understand domain-specific terminology~\cite{novick2009micro, kiani2019beyond, lee2010usability}. Compounding these issues is a lack of built-in structured guidance, which pushes users toward external resources, such as video tutorials, forums, and increasingly, large language models (LLMs), to navigate the UI and meet their task and learning goals. The continual toggling between disparate help sources disrupts workflow and frequently fails to deliver contextually appropriate support, thereby hindering self-directed learning and exploration~\cite{rieman1996field, chilana2012lemonaid, chilana2018supporting, khurana2024and}.

Multiple approaches have attempted to mitigate these challenges. One common strategy has been to provide simplified software alternatives, such as dedicated beginner modes~\cite{mcgrenere2002evaluation}, separate novice-oriented applications with simplified functionality~\cite{tinkercad, illustratoripad}, or AI-generated task-specific interfaces~\cite{vaithilingam2024dynavis}. While these alternatives reduce initial complexity, they often embody a simplicity-power tradeoff~\cite{uiandvis1999hearst}, where users may plateau in simplified environments~\cite{cockburn2014supporting}, or face a context switch when transitioning to the professional suite, which negates prior learning~\cite{nicol2024psychology, li2022design}.

Another line of work has focused on improving in-context assistance. Efforts range from static tooltips~\cite{grossman2010toolclips} and context-sensitive help~\cite{kelleher2005stencils} to more sophisticated systems that embed external resources directly into the application context~\cite{zhong2021helpviz}. However, these approaches struggle with accurately inferring user goals, potentially leading to intrusive or irrelevant help~\cite{brdnik2022intelligent}, and users may still find it challenging to map the provided information onto the complex software interface~\cite{kosch2023survey}. 

More recently, the advancement of LLMs has led to the emergence of in-application software copilots that can process natural language queries, monitor visual context, or automate tasks~\cite{firefly, krol2025copilotvision, khurana2025me}. While powerful for automating well-defined operations, current copilots struggle with complex, creative, or exploratory tasks~\cite{khurana2024and, qiao2024benchmarking}. Furthermore, prior work has found that excessive automation or guidance by these copilots can reduce engagement~\cite{kosmyna2025your}, diminish agency~\cite{li2024user, sellen2024rise}, reduce the user's ability to learn underlying concepts~\cite{chen2023understanding, vorvoreanu2025fostering}, and contradict the well-established maxim of \textit{learning by doing}~\cite{carroll1984training, rieman1996field}. Lastly, excessive automation and prescriptive guidance from copilots can lead users to adopt a passive, instruction-following mode, akin to the studied ``GPS effect''~\cite{ruginski2019gps}, instead of critical thinking and learning~\cite{swanson2026aifluency}.

Despite substantial prior work, existing systems still fall short of holistically reducing UI complexity, offering in-context structured guidance, and helping users interpret domain-specific terminology in ways that integrate learning into task execution. Motivated by instructional-scaffolding research where temporary support helps learners build competence and gradually work independently~\cite{belland2017instructional}, we introduce \textbf{\papername{}}, \textbf{a method that generates task-conditioned scaffolded interfaces to support software educators and tool creators in addressing the learner challenges} using the underlying native software's UI elements and operations. As shown in Figure~\ref{fig:teaser}, we demonstrate our LLM-based approach to generatively surface task-relevant tools from the underlying software and present them in a simplified layout tailored to the user's current task. The resulting scaffolded UI progressively expands in complexity as the user gains proficiency, links interface elements with domain concepts to support concept learning, and guides the user through task workflows while remaining grounded in the native software environment.

We apply our method to build two task-conditioned scaffolded interfaces in Blender\footnote{\url{https://www.blender.org/}}, a professional open-source 3D creation environment used as our testbed. We evaluate the two interfaces, one for 3D modeling and one for animation, with beginner (N=32) and expert (N=8) users. 
From the study with beginners (the software learners), we found that our scaffolded interfaces significantly reduced user-perceived task load, improved task performance through embedded workflow guidance, and increased domain concept understanding and learning in Blender during task execution. From the study with experts (the advanced users who have developed skills and knowledge through learning), we observed improved task efficiency and uncovered the potential to create personalized UIs for productivity and creativity needs of expert users. 
Based on a discussion of these findings, we present insights for future research on using our method to create scaffolded interfaces for instructional use cases to support both educators and learners, and to allow personalized workflows and cross-platform synchronization for end users. In summary, this work makes the following contributions: 

\begin{itemize}[leftmargin=*]
    \item We propose a \textbf{task-conditioned scaffolded interface generation method} to reduce UI complexity of professional creative software, streamline task guidance through UI configuration, and augment domain concept learning during task execution. 
    \item We present a \textbf{technical pipeline that applies the method} to generate scaffolded interfaces in professional creative software based on LLM-assisted UI generation. 
    \item We conduct a \textbf{user study} with both novices and experts to demonstrate the effectiveness of the scaffolded interface generation method for task execution and concept learning support. 
\end{itemize}

\section{Related Work} \label{sec:relatedwork}
\begin{figure}
    \centering
    \includegraphics[width=\linewidth]{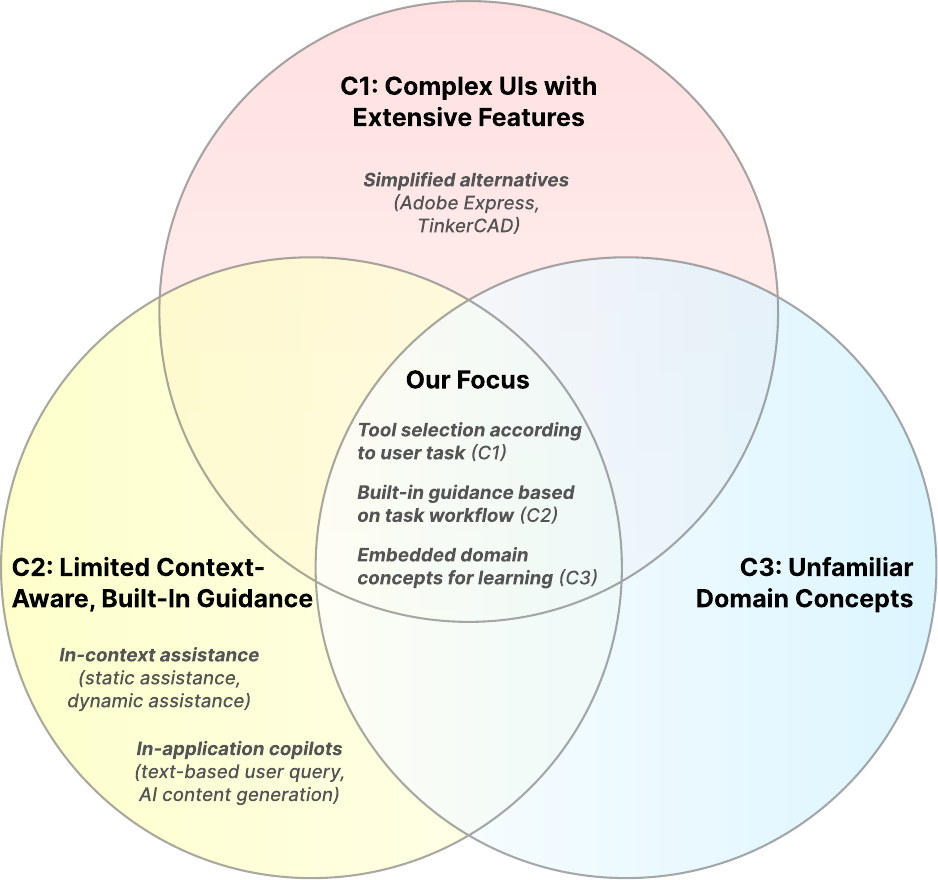}
    \caption{Challenges in professional creative software (C1--C3), prior research to tackle these challenges, and our focus.}
    \Description{This figure shows a venn diagram. There are three circles, each representing one of the three challenges this work addresses: C1 is complex UIs with extensive features, C2 is limited context-aware, built-in guidance, and C3 is unfamiliar domain concepts.}
    \label{fig:related_work}
\end{figure}

Our work builds upon prior research to address learning-related challenges in professional creative software. Figure~\ref{fig:related_work} presents a visual summary. 

\textbf{Challenges in professional creative software.}
Professional creative software has long been recognized as having a steep learning curve. A primary challenge is the complexity of the interface, which presents numerous icons, tool palettes, and menus~\cite{chilana2018supporting}. Prior work has identified that beginners frequently have trouble finding tools and determining the operational sequence for their tasks~\cite{novick2009micro, lee2010usability, kiani2019beyond}. 
Another challenge is that the software provides limited guidance on how to proceed with user tasks. To address these challenges, users often turn to learning resources which are mainly external, e.g., formal manuals~\cite{novick2006don}, tutorial videos~\cite{lafreniere2013community}, blogs, and dedicated forums, which disrupt task workflows and create a high barrier to self-taught exploration~\cite{rieman1996field, chilana2012lemonaid, chilana2018supporting, khurana2024and}. 
Additionally, professional creative software is designed for domain experts and its interfaces are often saturated with complex domain terminology, which raises  difficulty and steepens the learning curve for beginners~\cite{popovic2000expert}. 
In summary, the challenges of professional creative software that obstruct the learning experience include \textbf{C1: complex UIs with extensive features}; \textbf{C2: limited context-aware, built-in guidance}; \textbf{C3: unfamiliar domain concepts}. 

\textbf{Simplified alternatives to professional creative software.}
To tackle these challenges, one approach has been to build software with simplified functionality for beginners, which addresses \textbf{C1}. For example, Adobe Express\footnote{\url{https://www.adobe.com/express/}}, Canva\footnote{\url{https://www.canva.com/}}, OpenShot\footnote{\url{https://www.openshot.org/}}, TinkerCAD\footnote{\url{https://www.tinkercad.com/}}, and Illustrator on iPad\footnote{\url{https://www.adobe.com/products/illustrator/ipad.html}} target newcomers with simpler UIs and basic operations. In addition, recent work has explored natural language interfaces~\cite{generativefill, qin2024instructvid2vid, poole2022dreamfusion, liu2024dancegen} or LLM-generated interfaces~\cite{vaithilingam2024dynavis, liu2025crowdgenui}, with simple interaction modalities like text or on-demand UI controls. 
These alternatives embody the classic simplicity-power tradeoff~\cite{uiandvis1999hearst}, where simplified interfaces are less intimidating for novices, but they often lack the breadth of functionality. To balance this tradeoff, prior research has explored the transition between multiple interface modes, such as multi-layer interfaces~\cite{shneiderman2002promoting, mcgrenere2002evaluation} and workflow documentation interfaces~\cite{grossman2010chronicle}. 

Nonetheless, the gap between simple and advanced interfaces may lead to users getting stuck in the beginner mode if the two modes are too different~\cite{cockburn2014supporting}, or getting overwhelmed if advancing to complex modes is poorly managed~\cite{nicol2024psychology}. Another issue is that context switching from novice to professional software requires users to relearn workflows because the interfaces differ. Such discontinuity can void some of the initial learning done in the simplified application~\cite{li2022design}. 
Our work aims to address these limitations by embedding a task-conditioned scaffolded interface into professional creative software to reinforce connection with the native software and reduce discontinuity from platform switching. 
To manage UI complexity and help users move from basic to advanced, we offer user-selectable complexity levels that let them gradually take on more advanced tools as they learn the underlying concepts and work through their tasks.

\textbf{In-context assistance for professional creative software.}
To address \textbf{C2}, research efforts have explored in-context assistance in professional creative software. Unlike external resources, such assistance aims to provide support at the user's point of need, thus reducing disruption. For instance, previous work has designed elements, such as tooltips or help buttons, that are linked to relevant sections of a manual~\cite{farkas1993role, kelleher2005stencils, grossman2010toolclips, fourney2014intertwine, photoshop}. While integrated into the application, these elements are static and often lack user context sensitivity. 
Recognizing the limitations of static help, researchers have developed context-aware approaches. One thread of work explores \textit{intelligent user assistance}, such as using Bayesian probability networks to model user goals and offer help without explicit user requests~\cite{li2005active, horvitz2013lumiere}. Another line of work focuses on improving \textit{contextual assistance delivery} by bringing external resources into the application context, such as automatically presenting relevant web pages, documentation sections, forums, or tutorial videos based on the user's current activity~\cite{chilana2012lemonaid, fraser2019replay, matejka2011ambient, matejka2011ip, ekstrand2011searching, zhong2021helpviz, yang2022softvideo, yang2024aqua}, or showing procedural documentation with suggested next steps based on multiple prior demonstrations~\cite{bergman2005docwizards}.

However, existing in-context assistance faces challenges in accurately inferring user goals and risks providing irrelevant or poorly timed assistance~\cite{brdnik2022intelligent}. In addition, while contextual assistance delivery reduces the need for explicit searching, users can still struggle to identify the most relevant information~\cite{lim2012improving} and map the provided guidance (textual steps or visual cues) onto the complex interfaces~\cite{kosch2023survey}. Graphstract~\cite{huang2007graphstract} addresses the guidance-interface correlation challenge by using screen snippets centered around interaction areas to offer a visual overview of tools needed for user tasks. Although such graphic guidance can allow users to stay focused on task-related UI interactions, the task completion workflow remains implicit unless a detailed textual instruction is provided. In addition, for tasks that require using tools that are nested in menus or hidden in keyboard shortcuts and right-clicks, graphic guidance can become cluttered and ineffective.
We tackle these challenges by building a task-aware scaffolded interface that surfaces the relevant tools, triggered by UI, keyboard, and mouse interactions, from the underlying software. Rather than asking users to hypothesize instructions and navigate a complex UI, our interface embeds the guidance by grouping tools into workflow stages to build a clear task-tool alignment.
Crucially, most software assistance focuses on helping users complete tasks. Learning fundamental concepts during task performance has not been explicitly supported. We bridge this gap by integrating concept explanations into task instructions to actively engage users in learning concepts during tasks.

\begin{figure*}[!ht]
    \centering
    \begin{subfigure}[b]{0.38\textwidth}
        \includegraphics[width=\textwidth, height=6cm]{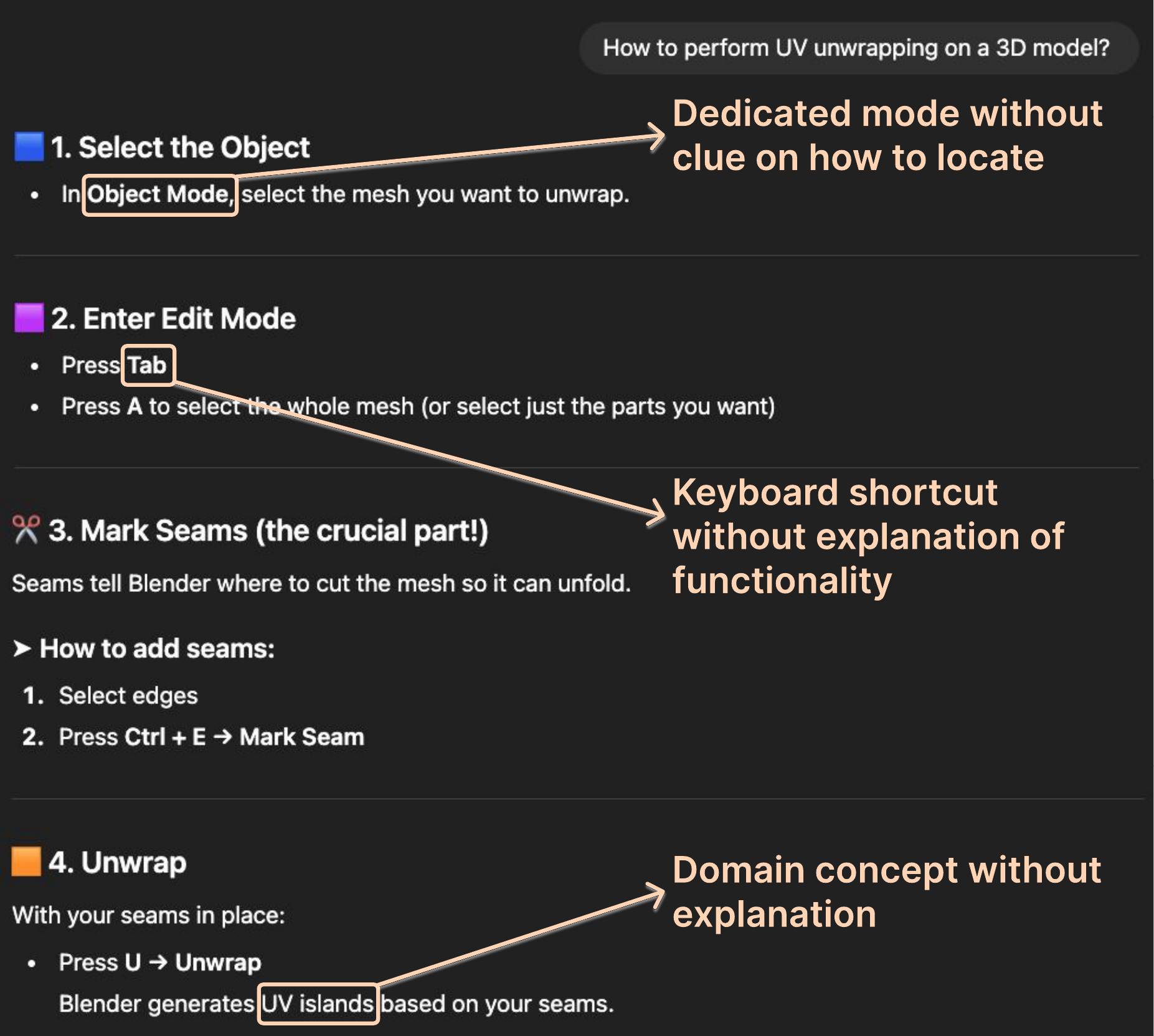}
    \end{subfigure}
    \hspace{0.5cm}
    \begin{subfigure}[b]{0.38\textwidth}
        \includegraphics[width=\textwidth, height=6cm]{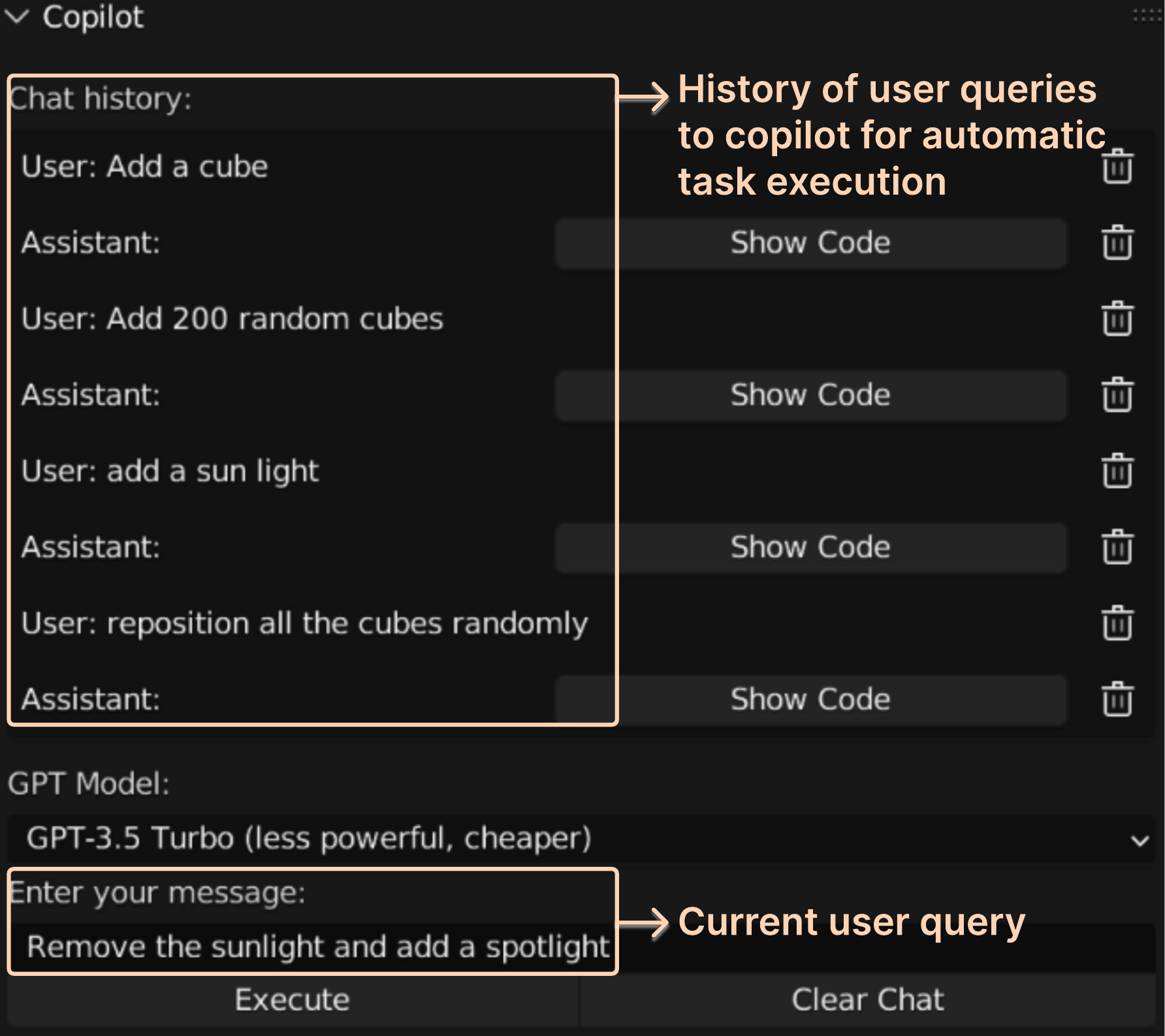}
    \end{subfigure}
    \caption{Examples of AI copilots that offer step-by-step task instructions~\cite{blender_copilot_oai} (left) or automatically execute a task~\cite{blender_copilot} (right) in Blender. The copilots' instructions overlook interface cues and operational context. By assuming familiarity with Blender's modes, shortcuts, and domain concepts, the guidance offers an incomplete procedural scaffold for beginner users.}
    \Description{This figure shows two examples of AI copilots in Blender. In the left figure, the copilot provides step-by-step task instructions, and in the right figure, the copilot automatically executes a user task.}
    \label{fig:copilot_example}
\end{figure*}

\textbf{In-application software copilots.}
The recent advancement of LLMs has reshaped the landscape of in-context assistance, leading to the emergence of in-application AI copilots. These copilots can process natural language queries, offer instructions, and automate software tasks, e.g., copilots in Blender as shown in Figure~\ref{fig:copilot_example}. These copilots can also use multimodal information, such as screen monitoring, to understand user context and offer guidance, e.g., Microsoft Copilot Vision~\cite{krol2025copilotvision, cunningham2025copilotvision} and Google AI Studio~\cite{google2025aistudio}. More recently, the copilots have unlocked the potential for on-demand UI generation, e.g., generative and malleable UIs~\cite{leviathangenerative, cheng2024biscuit, cao2025generative, min2025malleable}, which could be used to generate dynamic UIs to assist tasks based on user context and needs.

Although these copilots are helpful to guide or automate tasks, excessive guidance and task automation can diminish the user's sense of ownership over the task~\cite{endsley2017here, heer2019agency, li2024user, sellen2024rise}, decrease user confidence during the task completion process~\cite{xu2025productive}, and reduce cognitive engagement in learning~\cite{kosmyna2025your}. These copilots surface foundational concepts only incidentally, hindering a user's conceptual understanding (\textbf{C3}) and running counter to the principle of \textit{learning by doing}~\cite{carroll1984training, rieman1996field, novick2009micro, kiani2019beyond}.
The copilots might also suggest tools or UI elements that do not exist in the software or are not directly available, making the learning experience even more challenging. Although on-demand UIs can be generated by copilots to assist task execution, prior research has found that constantly changing UIs can increase cognitive load for the user, especially during longer task sessions~\cite{vaithilingam2024dynavis}.
To address these challenges and support durable learning, we embed explicit instructional scaffolds into the UI, aiming to engage the user in tasks. We bind each task-related action to its underlying concept to support learning while executing tasks. Furthermore, to reduce the cognitive load caused by dynamic UIs which may hinder learning, we offer users ready-to-use scaffolded UIs tailored to their tasks.

\section{\papername{}} \label{sec:system}
In this section, we present our method and implementation to generate task-conditioned scaffolded interfaces in professional creative software to support software learners by surfacing relevant tools, embedding domain concepts, and mapping the scaffolded interface back to the native software. The method is grounded in principles of task-centric design, adaptive interfaces, and scaffolded learning, and is implemented as a multi-stage pipeline based on LLM-assisted UI generation.

\subsection{Method}
\hspace*{1em} \textbf{Task-specific tool mapping.}
The method begins by extracting a set of tools required for the user's task. Hidden or nested functions are surfaced to reduce learning barriers and interface complexity (\textbf{C1}). This is motivated by task-centric UI design principles~\cite{lafreniere2014task} to manage interface complexity and reduce user cognitive load~\cite{sweller2011cognitive}. 

\textbf{Scaffolded interface construction.}
The method constructs an interface that organizes selected tools into workflow stages to offer task guidance and eliminate the need for users to switch between external tutorials and the software interface (\textbf{C2}). This is grounded in adaptive UI design principles~\cite{knowledge_navigator, jameson2007adaptive} to offer in-context, task-aware guidance through UI configuration. 

\textbf{Progressive disclosure.}
The scaffolded interface supports user-selectable levels of complexity, which enables beginners to start with fundamental functions and expand the interface as they gain proficiency. This aligns with the training-wheels interface model~\cite{carroll1984training} and intermodal guidelines~\cite{cockburn2014supporting} to manage the learning process and encourage exploration with user control.

\textbf{Concept labeling.}
Each tool is labeled with domain terminology to connect actions, concepts, and workflow steps, supporting concept learning during task execution (\textbf{C3}). This is motivated by prior research on learning by doing~\cite{rieman1996field, novick2009micro} and vocabulary extension~\cite{cockburn2014supporting} to explicitly reinforce learning engagement while performing tasks.

\textbf{Native-UI mapping.}
To prevent over-reliance on the scaffolded UI, each tool is linked to its location in the native environment through shortcut labels, menu paths, and interaction cues, which facilitates gradual transition toward native software fluency.

\subsection{Implementation}
\subsubsection{Technical Pipeline}
Here we present a technical pipeline to implement the task-conditioned scaffolded interface. In this work, we demonstrate an example that implements the scaffolded interface in Blender (version 3.6), a professional 3D creative software tool. The implementation pipeline involves using an LLM to assist with user task analysis, task-relevant tool selection, and UI code generation. This section outlines the technical pipeline and the complete implementation details for reproducibility can be found in Appendix~\ref{sec:llm_prompt}. Figure~\ref{fig:implementation} shows this pipeline. 

\begin{figure*}[!ht]
    \centering
    \includegraphics[width=\textwidth]{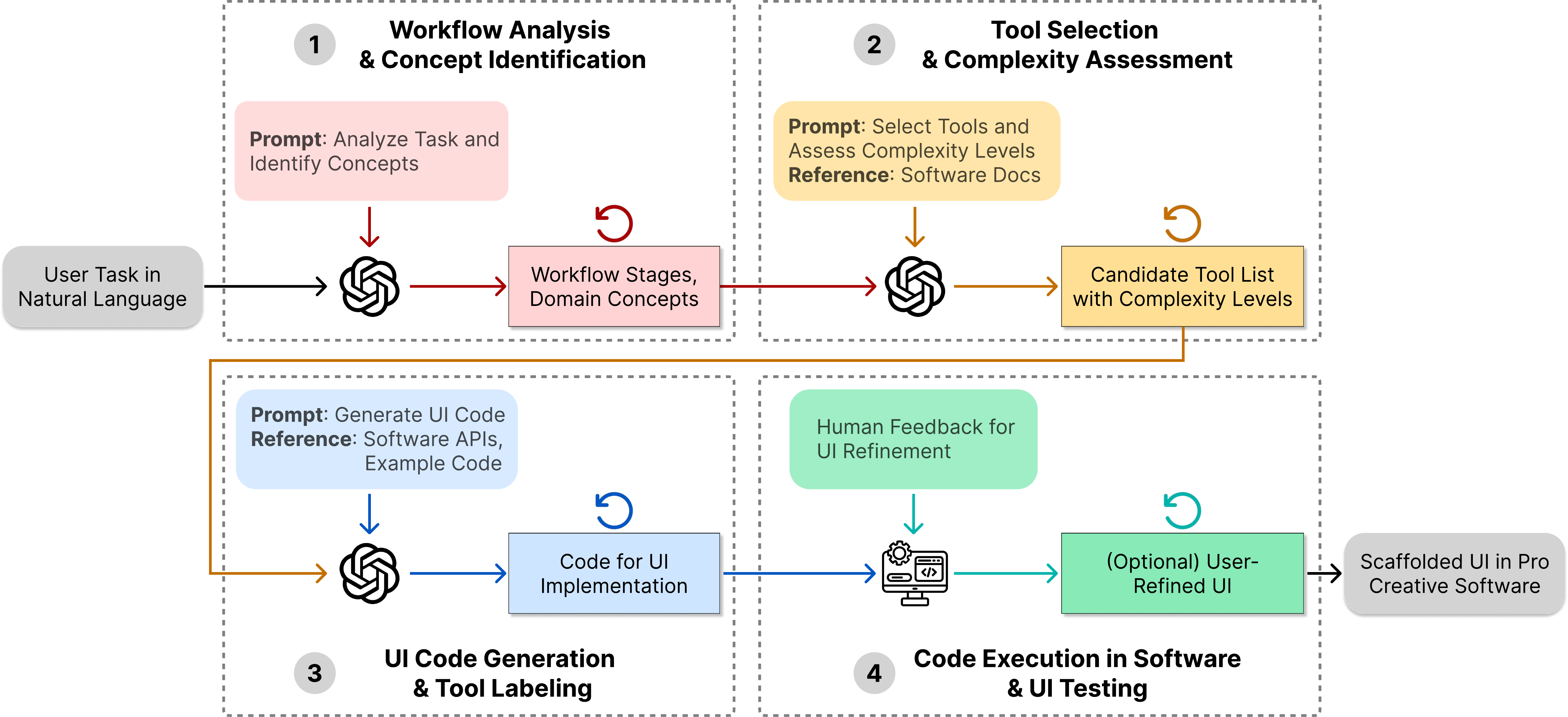}
    \caption{Technical implementation pipeline built upon the \papername{} method. (1) The pipeline takes a user's task description in natural language (e.g., perform UV unwrapping) that informs LLM-assisted task analysis to reason \textbf{workflow stages and relevant domain concepts} (e.g., marking seams, unwrapping, checking and visualization). (2) These workflow stages subsequently guide LLM-supported \textbf{tool selection and complexity assessment}, resulting in a candidate tool list (e.g., tools to mark seams and unwrap UV map). (3) The tool list is fed into LLM-based \textbf{code generation} for the UI implementation. (4) The code is executed to generate a \textbf{scaffolded interface} in professional creative software (e.g., custom panel in Blender). Users (e.g., UI designers, software educators, end users) are allowed to step in to refine LLM-generated artifacts (e.g., workflow, tool list, code) to adjust the scaffolded interface.}
    \Description{This figure shows the technical pipeline. Starting from the left, the pipeline input is the user's task, and it's passed to the LLM to reason the task workflow. The workflow is then passed to the LLM to select relevant tools. The tool list is further fed to the LLM to generate code for UI implementation.}
    \label{fig:implementation}
\end{figure*}

\textbf{Workflow analysis and concept identification.} 
The technical pipeline starts with taking a user's task description in natural language as the input. \papername{} then assembles the \textbf{workflow analysis and concept identification prompt} (Appendix~\ref{sec:llm_prompt_workflow}) that guides the LLM to decompose the user task into sub-tasks, forming the task workflow, and identify relevant domain concepts, which are associated with each sub-task. In this process, the LLM needs to analyze the required steps to perform the user task and classify these steps into workflow stages, as well as extract relevant domain concepts and explain them in a straightforward, concise description. 

\textbf{Tool selection and complexity assessment.} 
With the workflow stages and relevant concepts, \papername{} selects needed tools for each stage and maps them to the software functionality. For each workflow stage, \papername{} uses the \textbf{tool selection prompt} with reference to the software manuals for the LLM to select tools, e.g., keyboard and mouse operations and UI interactions. As part of this process, \papername{} assembles the \textbf{complexity assessment prompt} to instruct the LLM to suggest a complexity level for each tool (e.g., basic, intermediate, advanced) based on its functionality from software manuals (e.g., manuals in~\cite{blendermanuals}), and add these complexity levels to the tool description. Prompts for tool selection and complexity assessment are available in Appendix~\ref{sec:llm_prompt_tool}.

\textbf{UI code generation and tool labeling.} 
With the selected tools, \papername{} generates the code to implement each tool's functionality and renders the tools in a scaffolded interface. Specifically, \papername{} first assembles the \textbf{functionality code generation prompt} for the LLM to generate the code that executes each tool's functionality by referring to the software APIs (e.g., software APIs in~\cite{blenderapidocs}). Essentially, each keyboard and mouse operation or UI interaction is translated into an executable functionally-equivalent code block. Next, \papername{} uses the \textbf{UI code generation prompt} to guide the LLM to generate the code that converts the functionality code execution into UI interaction, e.g., button click or drop-down selection, such that users can initiate the execution of each tool through the scaffolded interface. In the UI code generation process, \papername{} provides the LLM with an example code that outlines the script's major components, including imported libraries, the interface layout (structures tools based on their corresponding workflow stages and groups tools according to their complexity levels), and main functions (link tools' functionality implementation to UI interaction). Based on our experiments, the example code can help to reduce syntax errors in LLM-generated code, such as missing imports, random UI layout, or inactive UI elements. Lastly, \papername{} assembles the \textbf{tool labeling prompt} to ask the LLM to add concept description as tooltips for each tool. Prompts for code generation are available in Appendix~\ref{sec:llm_prompt_code}.

\textbf{Code execution and UI testing.} 
In the last stage, \papername{} executes the generated UI code in the professional creative software to display the scaffolded interface (e.g., a custom panel in Blender). With this interface, UI designers, software educators, and end users, can test and evaluate the UI or make optional adjustments. They can adjust the intermediate artifacts offered to them in text, including the workflow stages, domain concepts, and tool selections, as well as the generated UI code for UI customization. Given the fact that LLM-generated outputs can be different from run to run even when the temperature parameter is set as zero~\cite{yuan2025understanding} and not always follow users' task instructions~\cite{zhou2023instruction}, the intermediate artifact inspection allows users to verify if the results are aligned with their requirements and make edits if needed. These user adjustments lead to iterative executions of the technical pipeline for the user to obtain updated scaffolded interfaces. Each execution takes 1-2 minutes. 

\subsubsection{Implementation Details}
We implemented our pipeline using GPT-4o as the LLM. The LLM-generated UI code occasionally contained errors such as missing imports or unbound UI controls. Providing example code as references helped to reduce these occurrences, and when errors did arise, the pipeline propagated them back to the LLM for self-correction, typically resolving within one to two iterations. As LLM coding capabilities advance, we envision increasingly robust pipeline execution.

\subsubsection{Generalizability} \label{sec:generalizability}
This technical pipeline can be used to accommodate different user tasks in the same and different software tools. Broader tasks are achievable by substituting a new task description, in natural language, as input to this pipeline. The pipeline will then analyze the task workflow, extract relevant domain concepts, select tools, and generate the UI code to create new scaffolded interfaces. Beyond a single software tool, this pipeline can be used to generate task-conditioned scaffolded interfaces in other software that provides a manual of its functionality for workflow analysis and tool selection, and programming APIs for UI code generation, such as UXP for Photoshop\footnote{\url{https://developer.adobe.com/photoshop/uxp/2022/}}, CUI for AutoCAD\footnote{\url{https://help.autodesk.com/view/OARX/2024/ENU/?guid=GUID-71554E76-8FD5-4853-82CD-3587764CBCAC}}, and OpenMayaUI for Maya\footnote{\url{https://help.autodesk.com/view/MAYAUL/2023/ENU/?guid=MAYA_API_REF_cpp_ref_group_open_maya_u_i_html}}. 
We further discuss the generalizability in Section~\ref{sec:discussion_generalizability} and demonstrate the implemented scaffolded UIs for variation tasks in Appendix~\ref{sec:interface_variations} and for other software (Maya) in Appendix~\ref{sec:interface_maya}.

\section{Illustrative Scenarios}
To illustrate the user experience in using our task-conditioned scaffolded interfaces, we walk readers through how a user can complete a 3D modeling task and learn relevant domain concepts. We compare the experience with performing the same task using the default interface of a professional creative software tool (Blender).

\textbf{Task introduction.}
UV unwrapping is a process for applying 2D textures to a 3D model's surface and requires domain concept understanding. It involves \textit{cutting} and \textit{unfolding} the 3D model into a flat 2D representation called a UV map. Each point on this 2D map, defined by U and V coordinates, corresponds to a vertex on the 3D model, allowing textures to be projected onto its surface. The major process includes defining seams---edges on the 3D model where the mesh will be split, and then executing an unwrap operation to generate the 2D layout of UV islands. Thereafter, these islands are arranged and optimized in the 2D space, and finally, the layout is checked to reduce distortions or overlaps. 

\begin{figure*}[!ht]
    \centering
    \includegraphics[width=\textwidth]{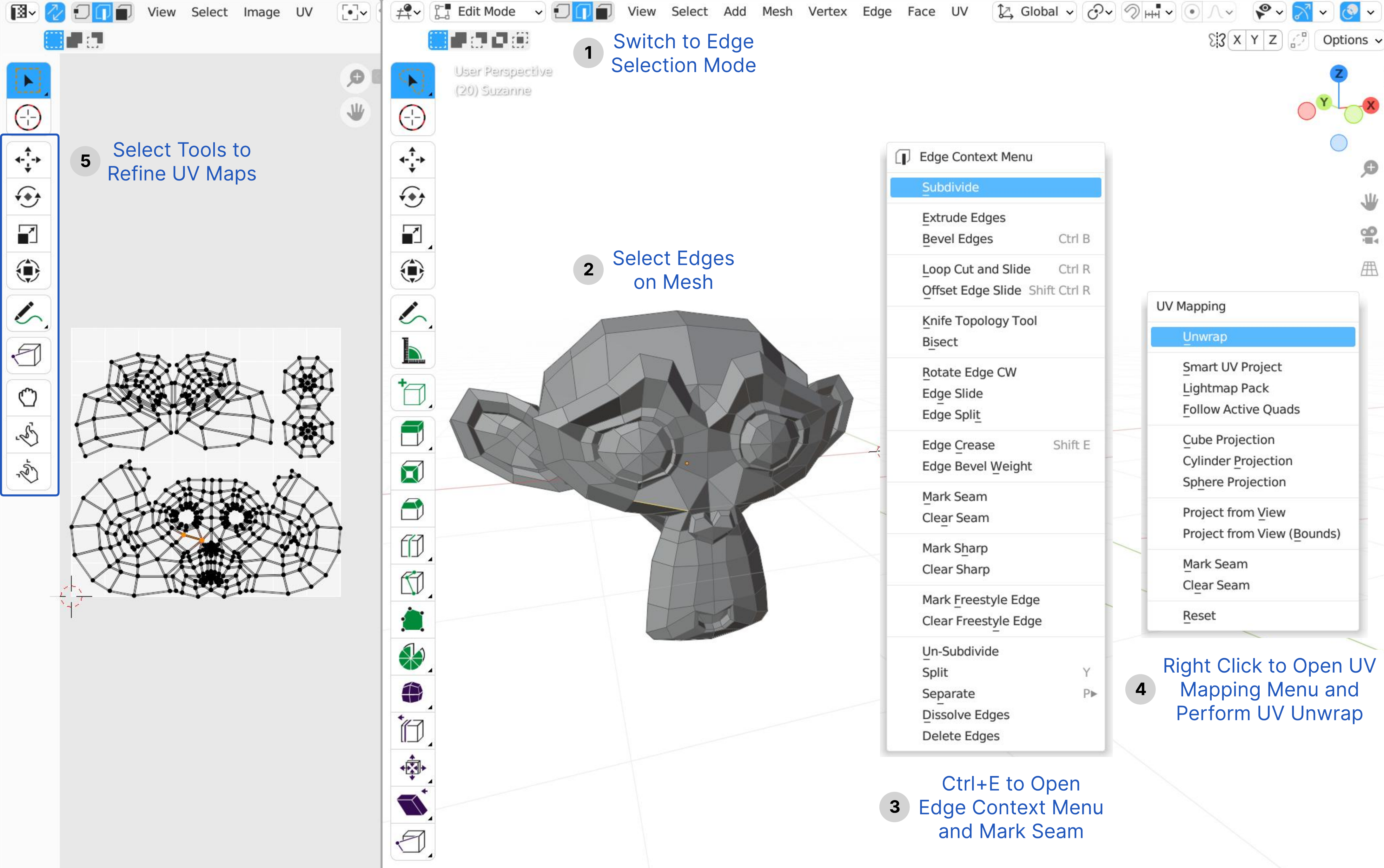}
    \caption{Default Blender UI to perform UV unwrapping. Tools are located sparsely and sometimes hidden in keyboard shortcuts or right-clicks. Domain concepts, e.g., ``seams'', ``unwrap'', and ``UV islands'', are opaque to beginners with limited explanation.}
    \Description{This figure shows the default Blender UI for UV unwrapping. Left is the UV editing region, which shows the UV map, and right is the 3D mesh operation region, which shows the 3D model.}
    \label{fig:use_scenario_default_ui}
\end{figure*}

\textbf{Performing the task using the default UI.}
Performing this task in the default software UI of Blender (Figure~\ref{fig:use_scenario_default_ui}) requires the user to navigate different menus and modes. The user enters ``Edit Mode'', switches to ``Edge Selection'' mode, and then starts selecting edges on the 3D mesh. After selecting the edges, the user marks them as seams in a dedicated UV menu via ``Ctrl+E''. Following it, they ``right click'' within the editing region to open an unwrapping operator to unwrap the 3D mesh along the marked seams. The resulting UVs are viewed and refined in a separate UV Editor using tools from a toolbar. 

During the process, the user spends significant amount of time following tutorials and familiarizing themselves with the operations and UI, such that they can locate and use tools through keyboard shortcuts, right clicks, and the UV Editor's toolbar to complete the task. In addition, domain concepts relevant to their task are not integrated into their task execution but instead requiring additional learning effort by searching online or taking courses.

\begin{figure}
    \centering
    \includegraphics[width=.85\linewidth]{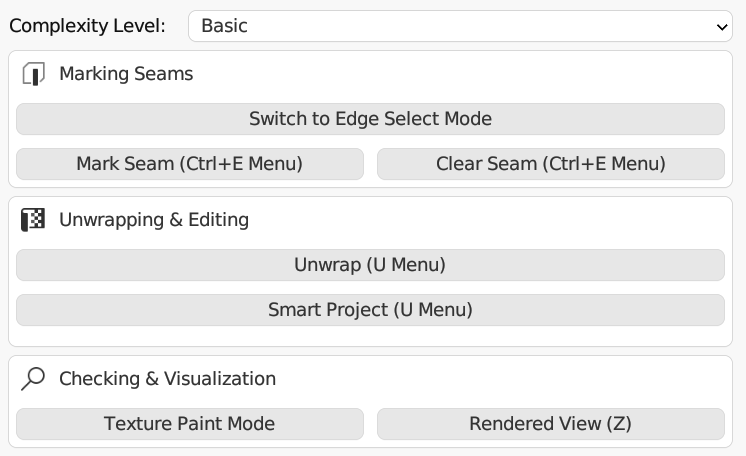}
    \caption{Our task-conditioned scaffolded UI's \textsc{Basic} level to perform UV unwrapping.}
    \Description{This figure shows the basic level interface. It displays the fundamental tools needed for UV unwrapping.}
    \label{fig:use_scenario_basic}
\end{figure}

\begin{figure}
    \centering
    \includegraphics[width=.85\linewidth]{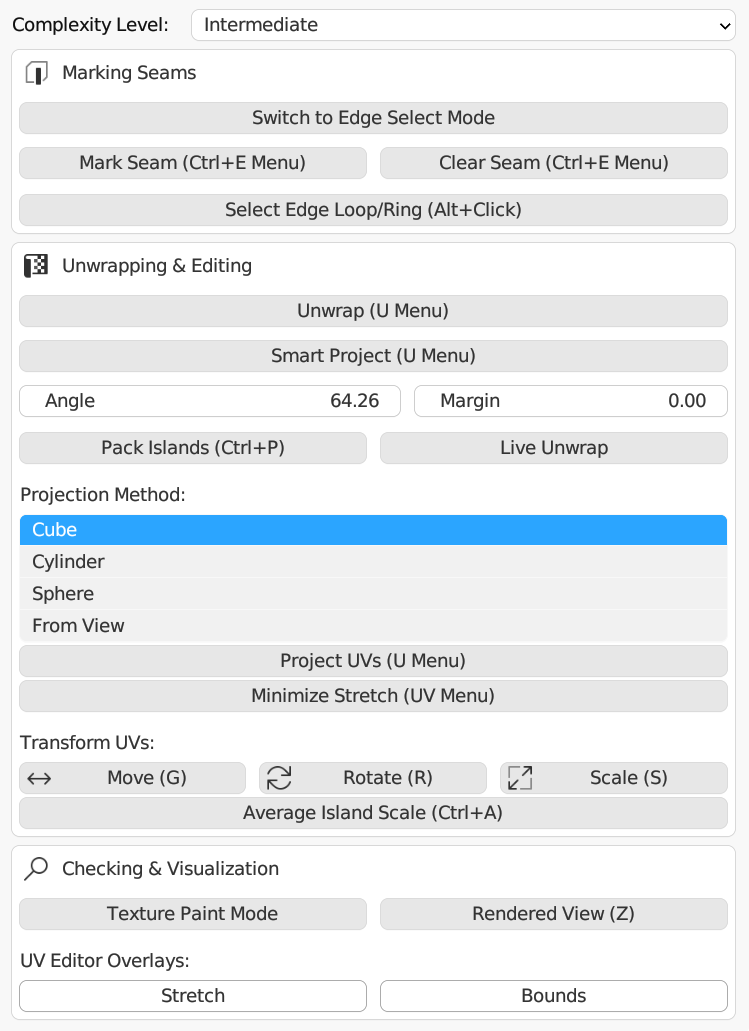}
    \caption{Our task-conditioned scaffolded UI's \textsc{Intermediate} level to perform UV unwrapping.}
    \Description{This figure shows the intermediate level interface. It displays the fundamental and some advanced tools that can be used for UV unwrapping.}
    \label{fig:use_scenario_intermediate}
\end{figure}

\begin{figure*}[!ht]
    \centering
    \includegraphics[width=\textwidth]{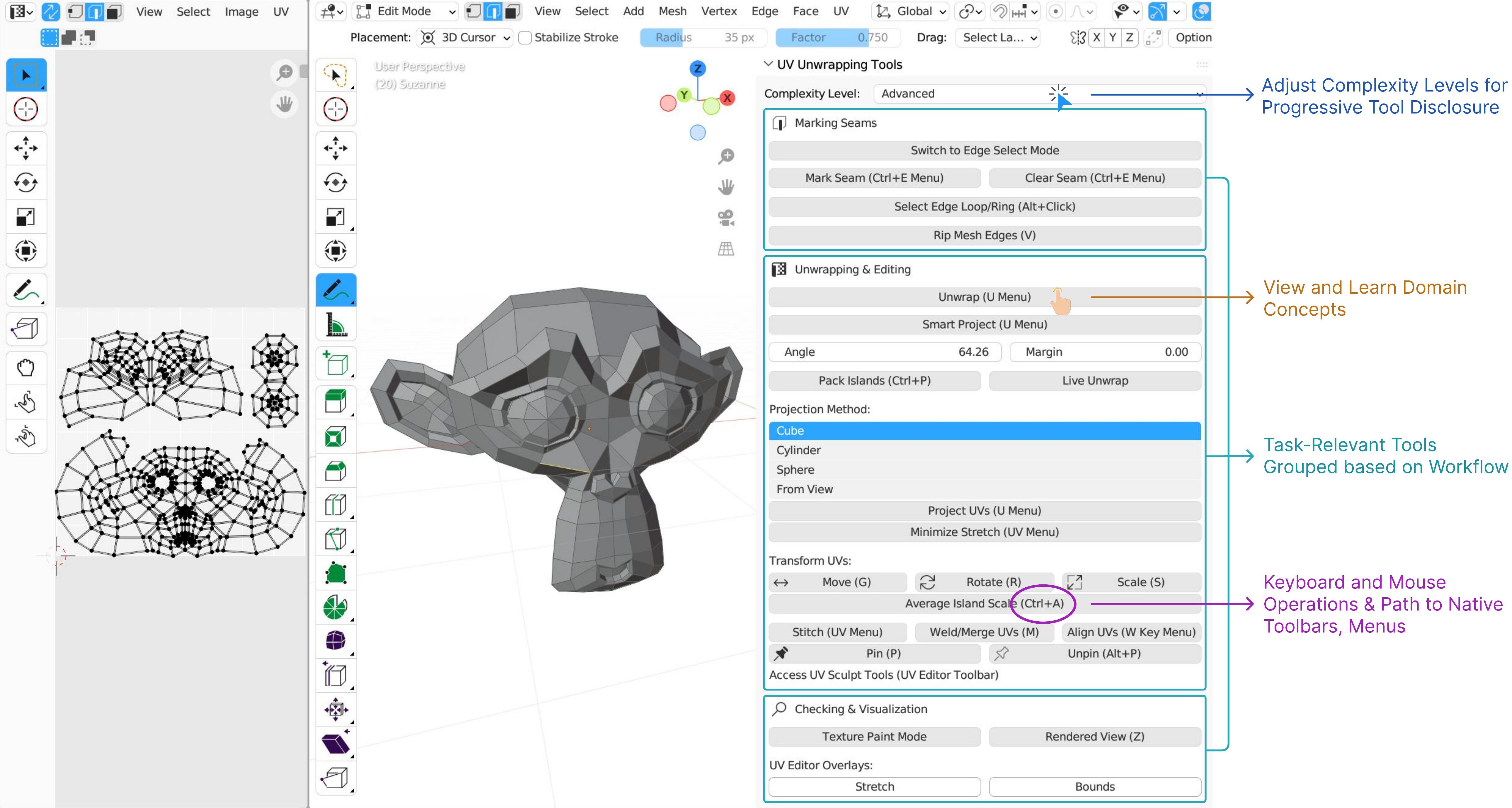}
    \caption{Our task-conditioned scaffolded UI generated to perform UV unwrapping. \textcolor{teal}{\textbf{Task-relevant tools}} are selected and grouped based on the workflow stages. Users can select complexity levels from basic, intermediate, to advanced for \textcolor{blue}{\textbf{progressive tool disclosure}}. Hovering over each tool shows a tooltip explaining the related \textcolor{orange}{\textbf{domain concepts}}. To connect the scaffolded interface with the \textcolor{violet}{\textbf{native software environment}}, keyboard shortcuts, mouse clicks, and the location of each tool in the menus or toolbars are labeled.}
    \Description{This figure shows the task-conditioned scaffolded UI for UV unwrapping. This scaffolded UI appears as an add-on panel on the right side in Blender's UI.}
    \label{fig:use_scenario_scaffold_ui}
\end{figure*}

\textbf{Performing the task using the task-conditioned scaffolded UI.}
Using our task-conditioned scaffolded interface, as shown in Figure~\ref{fig:use_scenario_scaffold_ui}, to work on this task, the user starts at the \textsc{Basic} complexity level (Figure~\ref{fig:use_scenario_basic}) and sees a structured task workflow. They first work through the \textit{Marking Seams} stage. All the tools needed to select edges and mark seams are organized under this stage. Following the suggested flow, they move to \textit{Unwrapping \& Editing} and use the relevant tools to unwrap the UV map and perform a simple visual check using tools in \textit{Checking \& Visualization}. 
Seeking to improve this initial result, they switch the interface's complexity level to \textsc{Intermediate} (Figure~\ref{fig:use_scenario_intermediate}) to utilize newly available seam selection aids and refine the UV layout. 
For final polishing and access to the complete toolset, they select the \textsc{Advanced} level (Figure~\ref{fig:use_scenario_scaffold_ui}) to reveal specialized tools. At this level, they can employ advanced operators for accurate adjustments and use analysis features to minimize distortion and verify the UV map quality. 

While working on the task, the user can reveal relevant domain concepts by hovering over each tool to see the tooltips. For instance, ``\textit{Mark Seam: Where to `cut' the 3D model's surface so it can be unfolded into a flat 2D layout}'', ``\textit{Unwrap: Flattens the 3D model's surface into 2D space based on the marked seams}'', and ``\textit{UV Islands: Makes UV islands (separate unwrapped parts) have the same relative size so textures don't look distorted}''. 
They can also see the keyboard shortcuts, mouse clicks, and native locations of each tool in its label to get familiar with the native software environment. 

During the process, the user is guided by the task workflow indicated by the interface without needing to switch back and forth between tutorials and their task. They can find task-relevant tools grouped under the corresponding workflow stage with less struggle remembering tool locations or dedicated keyboard shortcuts and mouse clicks. When working on task, they can control the tool complexity from basic to advanced based on their expertise and progress. They also get the chance to link their operations to the relevant domain concepts and learn these concepts during task execution. Although the scaffolded UI is shown on a separate custom panel, they can connect each tool with the default software's menu, keyboard shortcuts, and mouse operations through tool labels. 

\section{User Study Design}
To evaluate our task-conditioned scaffolded interfaces, here we introduce the user study design with beginners, who are the software learners (Study~1; Section~\ref{sec:study1_design}), and with experts, who are advanced users developed their skills and knowledge through learning (Study~2; Section~\ref{sec:study2_design}).

\subsection{Study 1: Beginner Users} \label{sec:study1_design}
From the study with beginners (the software learners), we explore the following research questions (RQs) to evaluate the effectiveness of the task-conditioned scaffolded interface compared to the baseline (default Blender interface) regarding task performance, concept learning, and software experience. 

\begin{itemize}[leftmargin=*]
    \item \textbf{RQ1: How does the task-conditioned scaffolded UI influence task performance?} This question explores whether task-aware interfaces with relevant tools impact user-perceived task load, task performance, task completion time (efficiency), and user confidence.

    \item \textbf{RQ2: How does the task-conditioned scaffolded UI affect the learning of domain concepts and software operations?} This question evaluates how progressive tool disclosure, concept-integrated UI organization, and bridging to native software help users understand domain concepts, connect concepts to tools, and learn the native software operations.

    \item \textbf{RQ3: How does the user experience differ with the task-conditioned scaffolded UI?} This question investigates the overall user experience, including user-perceived ease of use, sense of support vs. constraint, entry barriers, and reliance on external help and documentation.
\end{itemize}

\textbf{Experiment design.}
We adopted a between-subjects design where each participant was randomly assigned to one of two interface conditions (default Blender interface or our task-conditioned scaffolded interface) and to one of two task conditions. Task 1 asked participants to perform UV unwrapping on a cube, a beginner-level texturing task, while Task 2 involved constructing a human walk cycle, a basic animation task. We chose these tasks because they represent two main task pipelines of 3D modeling and animation, and require domain concept understanding. These tasks also accommodate adjustable complexity levels for us to evaluate with both beginner and expert users (Section~\ref{sec:study2_design}).
We provided all participants with the same cube for Task 1 and the same rigged human character ready for animation for Task 2 to ensure experiment consistency. The interfaces for Tasks 1 and 2 are shown in Figure~\ref{fig:use_scenario_scaffold_ui} and Appendix~\ref{sec:use_scenario_s2_scaffold_ui}, respectively. 
To ensure consistent study instruction, we provided a study guide (Appendix~\ref{sec:study_guide}) that introduced the basics of UV unwrapping and animation, as well as publicly available tutorial videos, with durations of 4.5 minutes for Task 1 and 7 minutes for Task 2. Task 2 was more complex than Task 1 due to the longer completion time and more required steps. 

\textbf{Participant recruitment.}
We recruited 32 English-speaking beginners (11 female, 19 male, 2 non-binary; mean age = 27.31, SD = 7.34) with no or limited 3D modeling and animation experience on Prolific\footnote{\url{https://www.prolific.com/}} (100\% approval rate). Participants were randomly assigned in a fully balanced design to one of two interface conditions, (1) Baseline (default Blender interface), or (2) Ours (task-conditioned scaffolded interface), to complete either Task 1 or Task 2, resulting in eight participants per task-interface combination.

\textbf{Measurements.} \label{sec:study1_measurements}
We used four questionnaires with 7-point Likert scale responses. 
For RQ1, we used the NASA Task Load Index (TLX)~\cite{hart2006nasa} and a custom task performance questionnaire. For RQ2 and RQ3, we used custom questionnaires on concept learning and user interface experience. These custom questionnaires were designed based on broadly used standard questionnaires and an instructional model, including Software Usability Measurement Inventory (SUMI)~\cite{kirakowski1996software}, Instructional Materials Motivation Survey (IMMS)~\cite{keller2010imms}, and ARCS model of motivation~\cite{keller2009motivational}. All the questionnaires are available in Appendix~\ref{sec:study1_questionnaires}. 

\textbf{Procedure.}
We invited eligible participants to a Zoom meeting. Before the study started, we emailed a consent form, which outlined the study goal, procedure, duration, and data collection. After providing informed verbal consent, participants filled out a pre-study demographics questionnaire ($\sim$5 minutes).
We then shared a study guide via a Google Slides link in Zoom chat, which included task instructions and tutorials. Once participants confirmed they had read the guide and accurately described the task, we shared our desktop screen, gave them remote control, and asked them to complete their assigned task. With consent, we recorded the desktop screen to record their task actions. Participants were encouraged to think aloud, and we took notes when observing them. Based on a pilot study with two beginners that helped us estimate task completion time, we gave the participants 20 minutes for Task 1 and 30 minutes for Task 2. 
After the task, participants were invited to complete the post-study questionnaires ($\sim$10 minutes). The study concluded with a payment of 10 USD to participants who worked on Task 1 for completing the $\sim$35 minute study and 15 USD to participants who worked on Task 2 for completing the $\sim$45 minute study. 

\textbf{Data analysis.}
To evaluate the effects of interface and task on participant responses, we conducted Type II two-way analyses of variance (ANOVA)~\cite{seabold2010statsmodels} with \textit{Interface} (Ours vs. Baseline) and \textit{Task} (Task 1 vs. Task 2) as between-subjects factors to examine the main and interaction effects on our measurements. Following significant ANOVA effects, we performed post-hoc pairwise comparisons using Bonferroni-adjusted t-tests~\cite{Vallat2018pingouin} to identify specific differences.

\subsection{Study 2: Expert Users} \label{sec:study2_design}
For the study with experts (the advanced users who have developed their knowledge and skills through learning), we kept \textbf{RQ1--RQ3} for a comparative analysis across beginner and expert users. In addition, two further research questions were formulated to explore our task-conditioned scaffolded UI's impact on skill acquisition and our method's generalizability.

\begin{itemize}[leftmargin=*]
    \item \textbf{RQ4: How can the task-conditioned scaffolded UI potentially shape long-term skill development?} This question invites expert perspectives on whether the task-conditioned scaffolded interface can help refine skills, adapt tools to new tasks, and strengthen concept understanding for solving new problems. 

    \item \textbf{RQ5: What is the potential impact of applying the task-conditioned scaffolded UI generation method to other software?} This question asks for insights from experts familiar with additional professional creative software beyond Blender on the benefits and challenges of using task-conditioned scaffolded interface in different platforms.
\end{itemize}

\textbf{Experiment design.}
Each participant was assigned to one of two task conditions. Task 1 involved performing UV unwrapping on a 3D monkey head (Figure~\ref{fig:use_scenario_scaffold_ui}), an advanced texturing task, while Task 2 required constructing a human walk cycle that reflected personality, an open-ended animation task. We provided the 3D model for Task 1 and the rigged human character for Task 2 for experiment consistency. 
Task assignment was based on participant prior experience: only those with 3D model texturing experience were assigned to Task 1, and only experienced animators were assigned to Task 2. This allowed us to gather feedback grounded in the participants' prior expertise to enable comparisons between their experience with the Blender interface and the task-conditioned scaffolded interface.

\textbf{Participant recruitment.}
We recruited eight experts (3 female, 5 male; mean age = 28.87, SD = 3.66) on Prolific, requiring at least one year of Blender experience and preferring additional proficiency in tools like Maya or 3ds Max. All other criteria for gender, age, language, and Prolific history matched those of Study 1. Participants (S2-P1 to S2-P8) were each assigned to either Task 1 or Task 2, and completed their task using the task-conditioned scaffolded interface, with four experts per task. Detailed demographics appear in Table~\ref{tab:study2_demographics}.

\begin{table*}[!ht]
    \centering
    \caption{Study 2 (expert user study) participant demographic information.}
    \Description{This table summarizes the demographics of expert users, including their IDs, YoE, and experience in 3D modeling, animation, and additional software.}
    \scalebox{0.95} {
    \begin{tabular}{ccccl}
    \toprule
        \textbf{ID} & \textbf{Year of Experience} & \textbf{3D Modeling} & \textbf{Animation} & \multicolumn{1}{c}{\textbf{Additional Software Experience}} \\
        \midrule
        S2-P1 & 2 & \cmark & \xmark & SketchUp \\
        \midrule
        S2-P2 & 5 & \cmark & \cmark & Maya, ZBrush, MotionBuilder \\
        \midrule
        S2-P3 & 3 & \cmark & \cmark & Maya \\
        \midrule
        S2-P4 & 3 & \cmark & \cmark & 3ds Max, Mixamo, Unity, Unreal Engine \\
        \midrule
        S2-P5 & 6 & \cmark & \cmark & Unity, Maya \\
        \midrule
        S2-P6 & 4 & \cmark & \cmark & Tinkercad, 3ds Max, Maya \\
        \midrule
        S2-P7 & 1 & \cmark & \xmark & FreeCAD, Tinkercad \\
        \midrule
        S2-P8 & 5 & \cmark & \cmark & Substance 3D, After Effects \\
    \bottomrule
    \end{tabular}
    }
    \label{tab:study2_demographics}
\end{table*}

\textbf{Measurements.}
To evaluate RQs 1, 2, and 3, we used the same questionnaires as Study 1 and adapted the items to preference-based questions: each item was rated on a 7-point Likert scale (1 = strongly prefer our interface, 7 = strongly prefer the baseline interface). Additionally, we collected qualitative feedback by asking participants to elaborate on their experience with our interface regarding task performance, concept learning, and interface experience. 
For RQs 4 and 5, we gathered user qualitative data regarding our task-conditioned scaffolded UI's impact on skill development and the potential of using the task-conditioned scaffolded UI in other software. Refer to Appendix~\ref{sec:study2_openended_questions} for the open-ended questions for qualitative data collection.

\textbf{Procedure.}
We invited eligible participants to our study via Zoom. The pre-study session involving the consent and pre-study questionnaire had the same procedure as Study 1. 
Following this, we offered a brief introduction to the user's task ($\sim$5 minutes). After participants verbally confirmed that they were clear about the task requirements, they used the task-conditioned scaffolded interface to work on the task for 15 minutes through remote desktop control. We recorded the desktop screen with participant consent and took notes on their actions. After the task, participants completed the post-study questionnaires and answered the open-ended questions ($\sim$10 minutes). The study concluded with a payment of 10 USD for completing the 30-minute study. 

\textbf{Data analysis.}
The qualitative data was analyzed using thematic analysis~\cite{braun2006using} with a deductive strategy. The authors reviewed the participant responses, coded the data to extract relevant information centered around the research questions, refined the codes, and categorized themes.

\subsection{Study Design Remarks}
Since our study with beginners aimed to explore their learning experience using our task-conditioned scaffolded UIs in a complex creative tool rather than making strong causal claims, not a large sample size was the main priority. We used a between-subjects design to avoid cross-condition contamination during learning and skill acquisition. In the study with experts, we prioritized qualitative depth over large-scale inference, and the eight users frequently interacted with Blender and were able to articulate pros and cons of our task-conditioned scaffolded interface clearly. Also, we observed signs of saturation during thematic analysis of the qualitative data.

\section{User Study Results}
In this section, we summarize the user study findings with beginner (Section~\ref{sec:s1_results}) and expert users (Section~\ref{sec:s2_results}). The findings are organized in response to our research questions (as introduced in Sections~\ref{sec:study1_design} and~\ref{sec:study2_design}). 

\subsection{Study 1: Beginner Users} \label{sec:s1_results}
\subsubsection{Questionnaire Responses}
\begin{figure*}[!ht]
    \centering
    \begin{subfigure}[b]{0.486\textwidth}
        \includegraphics[width=\textwidth]{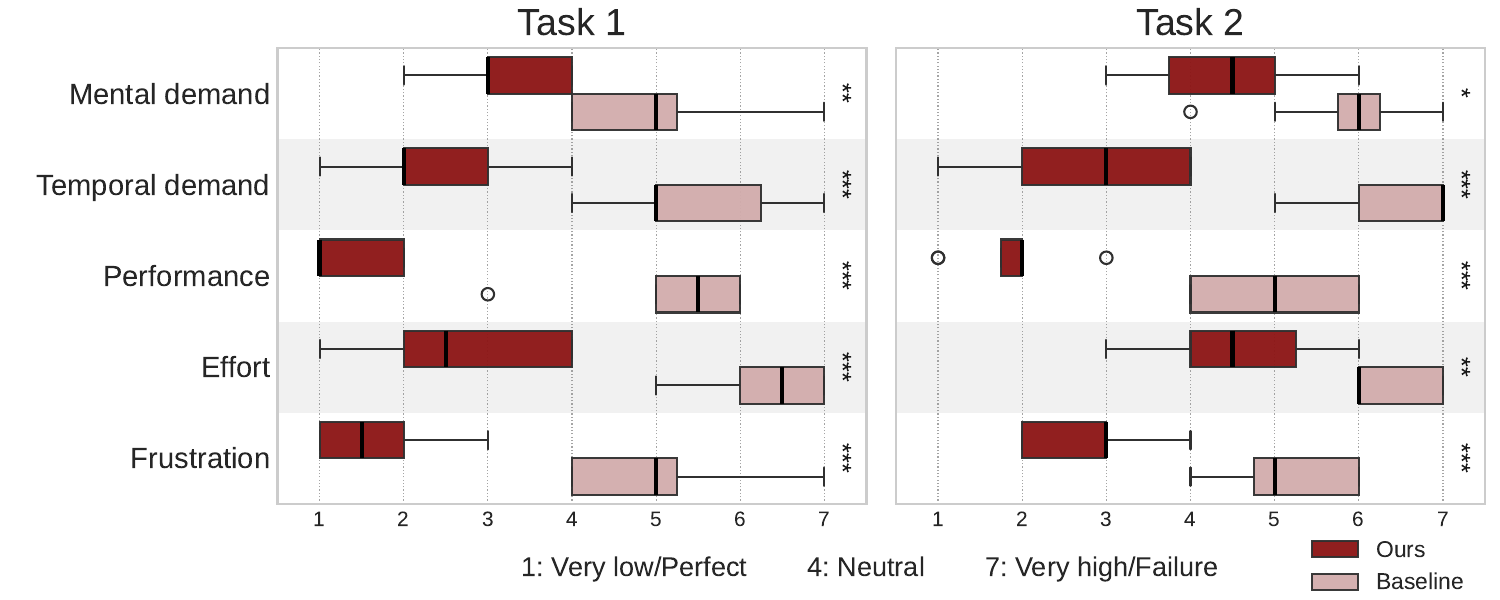}
        \caption{Task load.}
        \label{fig:study1_cognitive_load}
    \end{subfigure}
    \hfill
    \begin{subfigure}[b]{0.486\textwidth}
        \includegraphics[width=\textwidth]{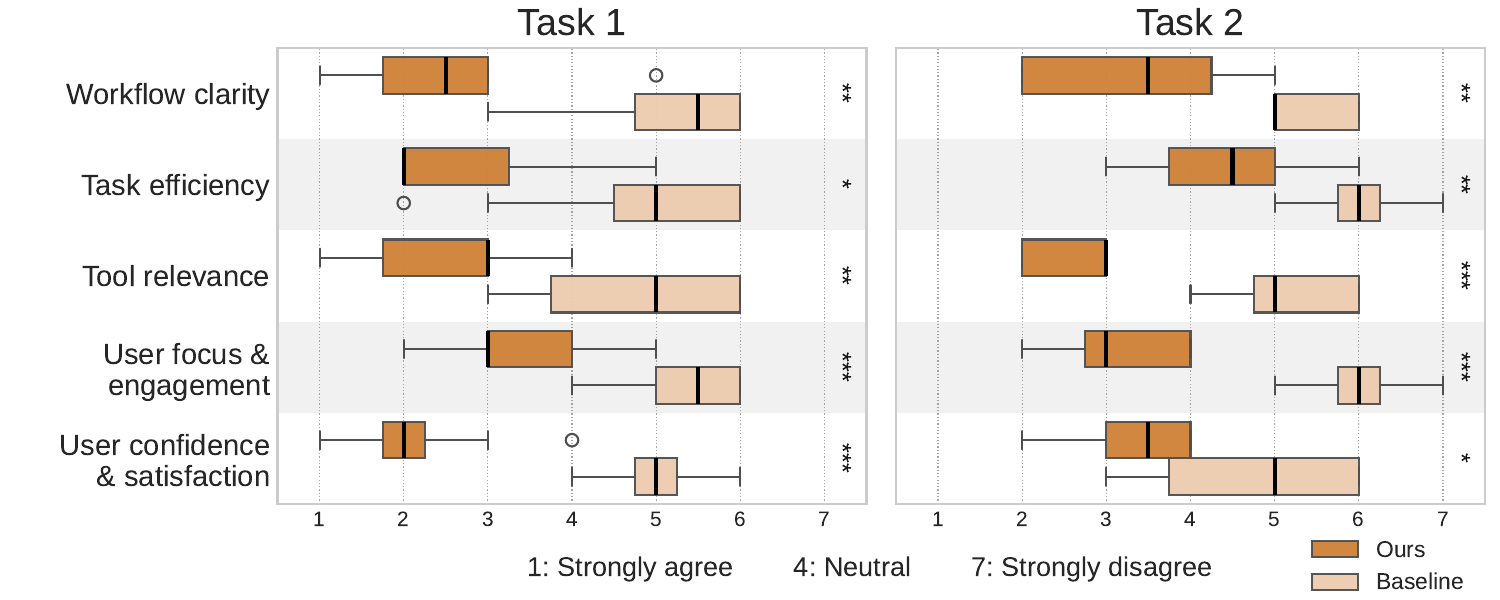}
        \caption{Task performance.}
        \label{fig:study1_task_performance}
    \end{subfigure}
    \hfill
    \begin{subfigure}[b]{0.486\textwidth}
        \includegraphics[width=\textwidth]{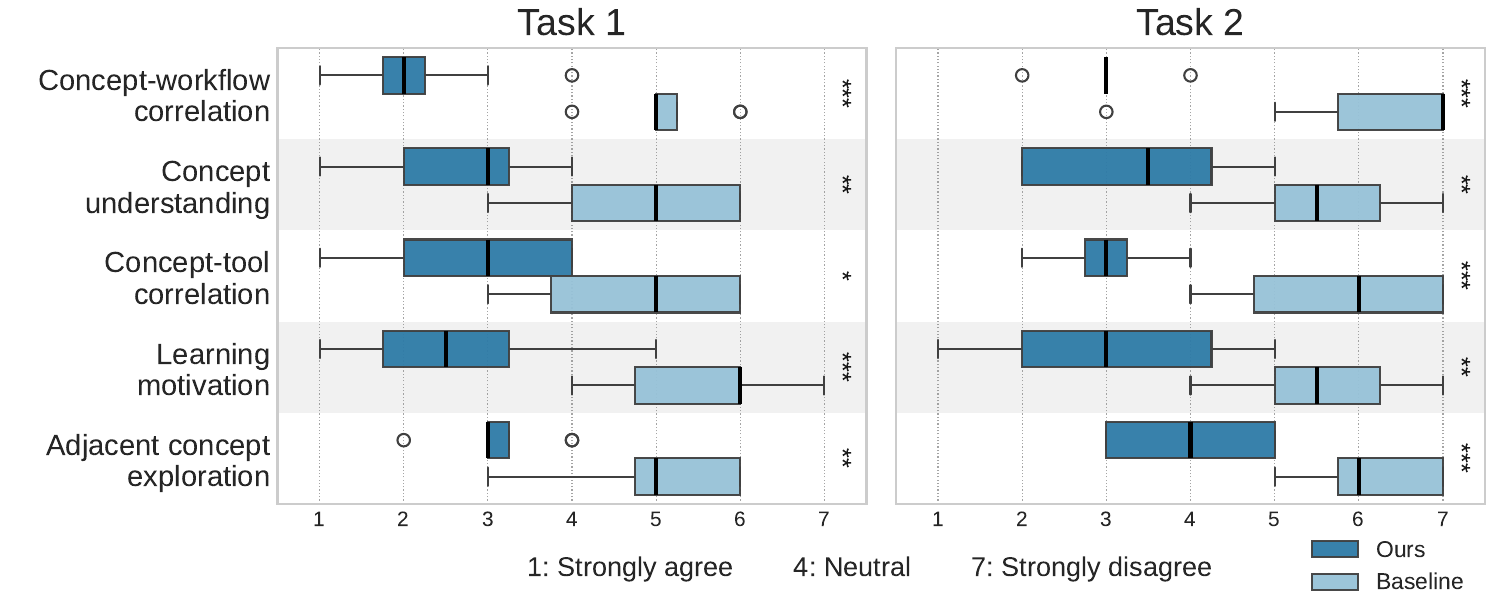}
        \caption{User-perceived concept understanding and learning.}
        \label{fig:study1_concept_learning}
    \end{subfigure}
    \hfill
    \begin{subfigure}[b]{0.486\textwidth}
        \includegraphics[width=\textwidth]{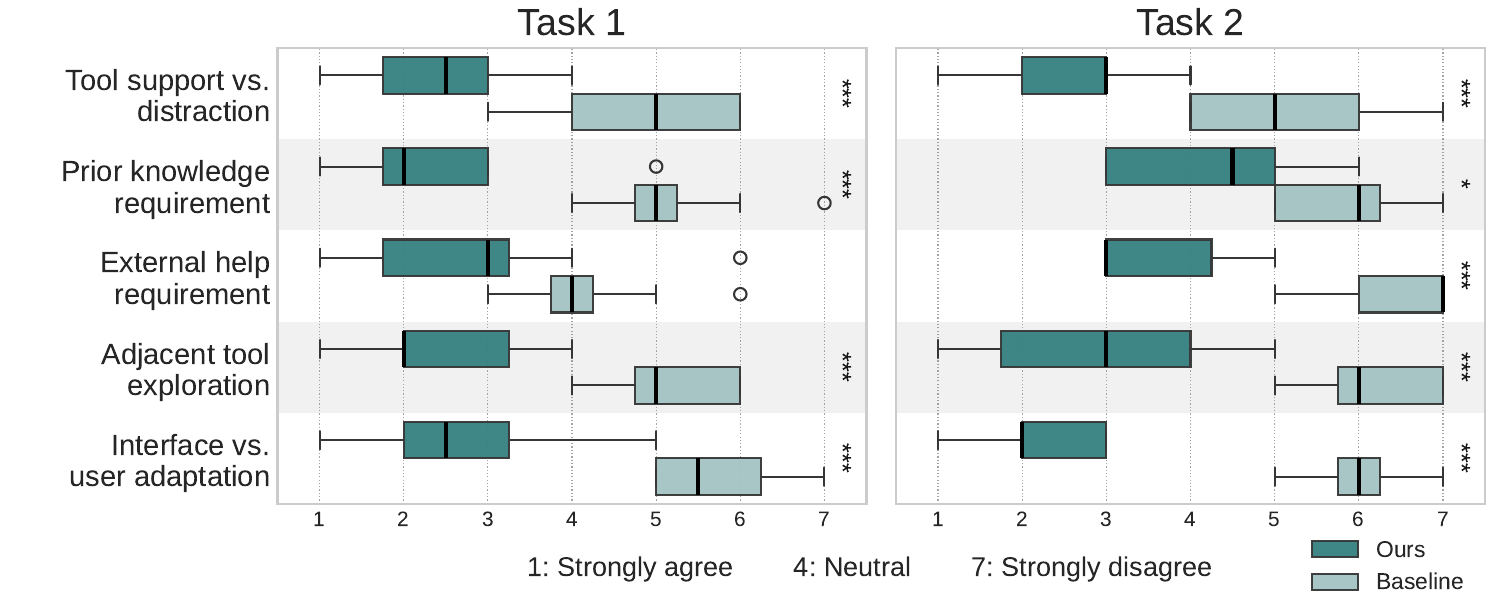}
        \caption{User interface experience.}
        \label{fig:study1_interface_experience}
    \end{subfigure}
    \caption{Study 1 (beginner user study) participant responses using Ours (task-conditioned scaffolded interface) and Baseline (default Blender interface) for Tasks 1 and 2 (lower is better). Asterisks denote significant differences in post-hoc pairwise comparisons between interfaces within each measure ($*$: $p < .05$, $**$: $p < .01$, and $***$: $p < .001$).}
    \Description{This figure shows the bar plots of Study 1 results. Our participants show clear preferences for our task-conditioned scaffolded interface across the measured four dimensions.}
    \label{fig:study1_measurements}
\end{figure*}

Figure~\ref{fig:study1_measurements} shows the participant reported responses to the questionnaires. Asterisks in the figure indicate statistical significance derived from post-hoc pairwise comparisons between Ours (task-conditioned scaffolded interface) and Baseline (default Blender interface) for each task and measure. The complete two-way ANOVA results are reported in Appendix~\ref{sec:study1_anova}. From the results, we see that our task-conditioned scaffolded interfaces have significantly improved task performance, supported user-perceived concept understanding and learning, and augmented user experience in using professional creative software compared to the default Blender interface. 

\subsubsection{Task Completion Time}
All participants successfully completed their tasks and generally spent less time using our task-conditioned scaffolded interface (Ours) compared to the default Blender interface (Baseline): 
\begin{align*}
    \bar{t}_{Ours (Task 1)} &= 12.69 \pm 1.49 \textrm{ mins}, \\ \bar{t}_{Baseline (Task 1)} &= 15.08 \pm 2.34 \textrm{ mins}; \\
    \bar{t}_{Ours (Task 2)} &= 20.39 \pm 2.55 \textrm{ mins}, \\ \bar{t}_{Baseline (Task 2)} &= 24.95 \pm 3.46 \textrm{ mins}.
\end{align*}
Significant main effects of \textit{Interface} ($F_{(1, 30)}=14.78, p<.001, \eta^{2}_{p}=.33$) and \textit{Task} ($F_{(1, 30)}=94.71, p<.001, \eta^{2}_{p}=.76$) were observed. The interaction effect was not significant. Post-hoc analysis revealed no significant difference in completion time in Task 1, but users with our interface were significantly faster in Task 2 ($p<.05$), demonstrating the efficiency advantage of our task-conditioned scaffolded interface under higher task demands.

\subsubsection{User Behavior}
We observed that participants using our task-conditioned scaffolded interface (experiment group) progressively reduced their reliance on tutorial videos and increasingly navigated via the interface's inherent workflow structure, echoed by participant self-reported improved workflow clarity, tool relevance, and reduced external help requirement. In contrast, participants using the default Blender interface (control group) consistently referred to the videos throughout the tasks. Consequently, the experiment group generally completed tasks faster than the control group. 
Additionally, when some tools were not mentioned in the tutorials but presented in our task-conditioned scaffolded interfaces, the experiment group explored them during or after their tasks, as echoed by participant self-reported tool and concept exploration. Conversely, the control group rarely explored beyond the tools covered in tutorials.

\subsection{Study 2: Expert Users} \label{sec:s2_results}
\subsubsection{Questionnaire Responses}
\begin{figure*}[!ht]
    \centering
    \includegraphics[width=0.85\textwidth]{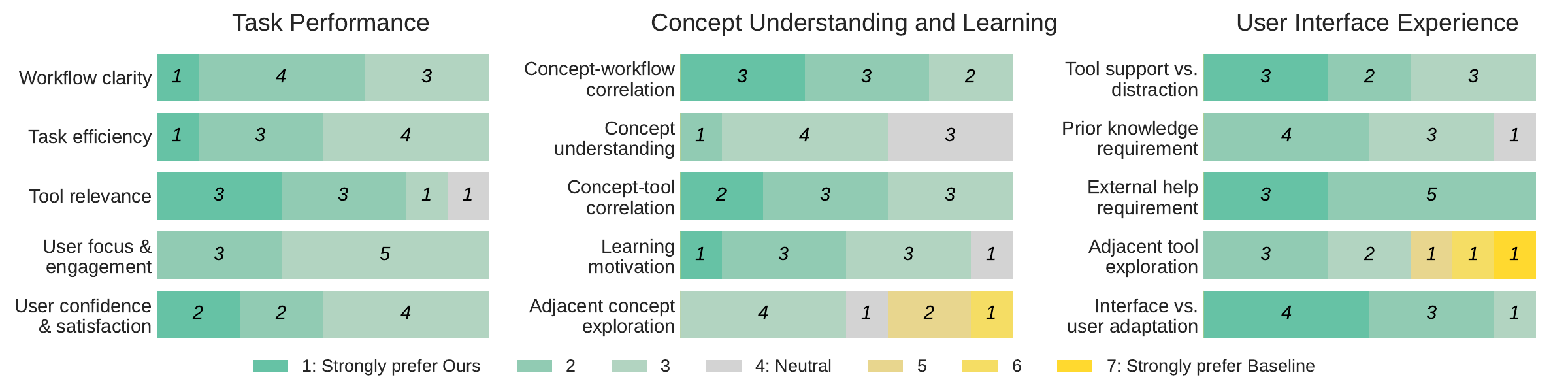}
    \caption{Study 2 (expert user study) participant preferences for Ours (task-conditioned scaffolded interface) or Baseline (default Blender interface) regarding task performance, user-perceived concept understanding and learning, and user interface experience.}
    \Description{This figure shows the bar plots of Study 2 results. For most questions from the three evaluated aspects, the participants preferred our task-conditioned interface, except for "Adjacent concept exploration" in perceived concept understanding and learning and "Adjacent tool exploration" in user interface experience, the participants had mixed opinions.}
    \label{fig:study2_scores}
\end{figure*}

Figure~\ref{fig:study2_scores} shows the participant reported preference scores for Ours (task-conditioned scaffolded interface) or Baseline (default Blender interface) on the measures of task performance, concept understanding and learning, and user interface experience. 
They had a clear preference for our interface across all dimensions regarding task performance. 
For concept understanding and learning, participants generally perceived our interface as more supportive for understanding concepts, relating tools to concepts and workflows, and motivating learning. However, they had mixed opinions on concept exploration (4 favored our interface, 1 was neutral, 3 favored the baseline).
For the overall interface experience, consistent preference for our interface was observed except for tool exploration (5 participants preferred our interface, 3 favored the baseline).

\subsubsection{Qualitative Results}
Here we present the themes derived from qualitative data we obtained from the expert users. 

\textbf{Tradeoffs of supporting task performance and allowing workflow flexibility (RQ1).}
Participants generally appreciated task performance support by surfacing task-relevant tools (S2-P1, S2-P2, S2-P7, S2-P8). The task-conditioned scaffolded interface led to higher task efficiency and user focus, as S2-P2 said that, ``\textit{I actually finished the UV task faster than expected.}''
However, participants also expressed concerns that the system's assumptions sometimes misaligned with their preferred workflows. Some found the interface limiting when it failed to surface user-expected tools (S2-P4). Others perceived the guided structure forced a specific process, which may reduce workflow flexibility (S2-P6). These findings indicate the need for more user customization to support expert users who prefer flexible or iterative workflows.

\textbf{Learning and exploration opportunities during task performance (RQ2).}
Participants liked our interface for its ability to clarify the task-related concepts, which can benefit educational or onboarding contexts (S2-P1, S2-P3). They also highlighted the learning benefits, saying that, ``\textit{it's not just clicking through menus but also teaching you why they matter}'' (S2-P2). 
In addition, progressive tool disclosure was appreciated as a way to guide structured concept learning. For instance, S2-P3 noted that, ``\textit{it shows graph editing after you're done with key posing, and this can help avoid overwhelming learners.}''
Furthermore, participants remarked that showing tools adjacent to the immediate task was helpful to motivate learning. As noted by S2-P5, ``\textit{It prompted me to notice things like secondary motion I'd normally skip past},'' indicating that the interface enables the discovery of related concepts and tools and encourages user exploration. 

\textbf{Gains of simplicity and focus vs. loss of familiarity (RQ3).}
Participants responded positively to the overall experience since our task-conditioned scaffolded interface was ``\textit{clean, not overwhelming}'' (S2-P1), ``\textit{task-centered}'' (S2-P4), ``\textit{lightweight and approachable}'' (S2-P7), and with ``\textit{limited distractions}'' (S2-P8). S2-P2 also noted that, ``\textit{I can imagine a less fighting the UI if I were a newbie},'' indicating reduced cognitive friction during software navigation for beginners. 
However, participants commented that the task-conditioned scaffolded interface may conflict with their familiarity with the default software layout. S2-P3 remarked that they ``\textit{missed the old familiarity},'' pointing to the tension when introducing novel interfaces. 

\textbf{Scaffolding's impact on skill development (RQ4).}
Participants found that the task-conditioned scaffolded interface provided a supportive foundation for building practical skills, particularly for newcomers. For example, S2-P5 called it a ``\textit{great stepping stone from beginner to intermediate},'' and S2-P6 noted the interface made it ``\textit{easier to build confidence in doing simple stuff}.'' These responses show that experts see our interface as helping beginners build confidence and gain skills while learning professional creative software.

\textbf{Using task-conditioned scaffolded interfaces in broader contexts (RQ5).} \label{sec:results_generalizability}
Participants with experience in other professional creative software recognized the underlying principles of task-conditioned scaffolded interface design as being transferable. For instance, S2-P8 stated that, ``\textit{I think the knowledge about motion, timing, bone constraints translates to using other tools}.'' 
Building on such transferability, participants suggested potential applications and impacts of using task-conditioned scaffolded interfaces in other software. S2-P6 said that, ``\textit{3ds Max can benefit from this. The modifier stack alone is complex enough. If tools were grouped by task and revealed based on what you're actually doing, it'd cut down on confusion}.'' Participants also anticipated the benefits, e.g., ``\textit{it could help new users get productive faster, and even experienced users might find it useful to stay focused}'' (S2-P6). These results suggest user enthusiasm for applying the task-conditioned scaffolded interface generation method more broadly to reduce interface complexity and enhance workflow efficiency.

\subsubsection{User Behavior}
We observed that expert users approached tasks flexibly, sometimes bypassing the step-by-step workflow stages in the task-conditioned scaffolded interface. For instance, S2-P6 switched between marking seams (stage 1) and unwrapping \& editing (stage 2). They started with the ``Select Edge Loops'' tool in stage 1 to mark seams, used ``Unwrap'' in stage 2 to check the UV map, then went back to use ``Edge Select'' in stage 1 to refine the seams and repeated the process. This reflects their preference for multitasking and iterative workflows. 
Furthermore, when needed tools were not presented at their current complexity level, several participants (S2-P2, S2-P5, S2-P6, S2-P8) sometimes did not look for these tools by advancing complexity levels. Instead, they accessed tools via keyboard shortcuts or the default Blender interface---the operations they were familiar with. This aligns with their feedback that the task-conditioned scaffolded interface was less supportive for tool exploration compared to the experience of beginners.

\section{Discussion}
Based on the user study findings, here we discuss the opportunities in \papername{}, implications for design, \papername{}'s limitations, and future directions.

\subsection{Opportunities in \papername{}}
\hspace*{1em} \textbf{Task-aware adaptation to support software learners and educators.}
In our study, beginner users reported benefits in both task performance and user-perceived concept understanding with our task-conditioned scaffolded interface, showing its promise for instructional use. To better support learning for software learners, the interface can adapt its scaffolding strategy based on task complexity. 
For simple tasks, the interface can serve as a guided playground where future design could expand the available tools slightly beyond the current step to promote curiosity-driven exploration. 
For complex tasks, the focus shifts to practice and learning, so the interface can offer a smaller, essential toolset with clear links between domain concepts and software operations. It could also include best-practice tips or visual aids like video clips to help ensure key concepts and operations are grasped before moving on. 

In addition, empowering educators is essential to unlock the full potential of using scaffolded interfaces for instruction. Educators could choose the concepts and tools the interface highlights, tailoring it to their teaching goals. 
Equally important is the control over the instruction process. If educators set when new tools and concepts appear, they can pace teaching based on student progress, creating a structured path that builds complexity gradually. 
By offering the blend of task-aware adaptation for learners and educators, scaffolded interfaces can evolve into helpful assistance for educational purposes. 

\textbf{Scaffolded interface driven by productivity and creativity.}
Although our work focused on building task-conditioned scaffolded UIs to support software learning for beginners, our study with experts revealed that advanced users valued productivity through personalized workflows and creativity by flexible exploration. Therefore, scaffolded interfaces targeting at expert end users should evolve beyond structured guidance towards systems driven by these needs.
To support productivity, a scaffolded interface could learn a user's unique shortcuts, tool sequences, and niche commands, and optimize workflows by rearranging palettes for top tools or suggesting more efficient alternatives.
Implementing this requires robust user activity logging, potentially through screen monitoring by in-application AI copilots~\cite{cunningham2025copilotvision, google2025aistudio}, to adapt interfaces according to the user's demonstrated behavior. 

To allow creative needs that step outside routine workflows, a scaffolded interface can act as a serendipity engine that encourages exploration. For instance, it can suggest unusual tool pairings based on deeper function relationships or highlight rarely used but relevant tools. This can be powered by LLMs, which use the user's current state and project context to generate uncommon operations or workflows. We explore this potential by implementing scaffolded interface variations adapted to multiple creative demands, as shown in Appendix~\ref{sec:interface_variations}. 

\textbf{Scaffolded interface meta-layer for cross-software workflows.} \label{sec:discussion_generalizability}
An enthusiasm by our expert users was using the task-conditioned scaffolded interface generation method in multiple software platforms. Experts highlighted the transferable task-aware, concept-driven UI design for task efficiency, using the scaffolded interface as a meta-layer to bridge different applications.
Imagine that end users could design their own scaffolded interface built upon the tools they frequently use. Instead of learning a new interface for every software, they could keep this consistent interface that maps familiar tools to different underlying software. 
Such an interface could help improve cross-software workflows as many complex projects require switching between multiple applications, and create a unified experience, reducing the effort required by switching contexts. In addition, this scaffolded interface meta-layer can help users stay focused on learning fundamental concepts and operations instead of spending significant efforts on learning each software's new interface. We demonstrate this possibility by implementing scaffolded UIs for our user study tasks (UV unwrapping and walk cycle animation) in an additional professional software tool, Maya, and the results are available in Appendix~\ref{sec:interface_maya}.

However, extending scaffolded UIs to broader domains and more varied tasks introduces foreseeable challenges. Tasks that are exploratory or open-ended in nature, e.g., freeform sculpting or experimental compositing, may resist the structured workflow decomposition that our current scaffolding relies on, and a constrained meta-layer could suppress the creative divergence these workflows demand. Additionally, maintaining semantic consistency across software platforms requires robust mappings between conceptually equivalent but operationally distinct tools, which may not always exist or may break as underlying applications evolve independently.

\subsection{Implications for Design}
\hspace*{1em} \textbf{Scaffolded UIs unify task instruction and interaction into a single design layer.} A fundamental distinction between scaffolded UIs embedded within professional creative software and in-context assistance mechanisms, such as screen snippets highlighting relevant UI elements~\cite{huang2007graphstract} or procedural documentation guiding users through workflows~\cite{bergman2005docwizards}, lies in how each approach structures the relationship between task instruction and UI interaction. In-context assistance treats the application's native interface and the instructional layer as separable concerns. Prior work has shown that with in-context assistance, users benefit most when visual guidance is tightly coupled with textual instructions, so that they can simultaneously understand \emph{what} to do and \emph{where} to do it within the interface~\cite{huang2007graphstract}. Yet even when well-integrated, these overlays remain fundamentally external to the underlying UI, i.e., they annotate it rather than restructure it. Our scaffolded UI, by contrast, abstracts a focused, task-specific interface from the full application, unifying instruction and interaction within a single layer. Because the scaffolded UI is a malleable design surface, it affords the embedding of task guidance, learning materials, and other potential content directly into the interaction, rather than layering them on top of an interface whose complexity may itself be a barrier, especially for beginner users.

\textbf{Tradeoffs between instructional integration and progressive skill transfer.} This unification eliminates the cognitive overhead of context-switching between task instruction and execution. In the meantime, it introduces a tradeoff, where the scaffold UI could constrain the user's exposure to the full application, while in-context assistance preserves the user's engagement with the native interface. These tensions are echoed by our expert participants' concerns about potential over-reliance on the scaffolded interface. Future design could include cues like UI minimaps or short animations showing where tools come from. The scaffolded interface could also fade as users gain skill, encouraging further transition to the native UI. Essentially, the design can adopt a transition from unified scaffolding to lightweight in-context cues, allowing the onboarding and learning benefits of scaffolding while ultimately fostering full interface fluency.

\subsection{Limitations and Future Directions}
We discuss our work's limitations here and offer insights for future research. 
First, our interface uses user-selectable complexity levels for progressive tool disclosure. This is just one point on the spectrum from user-driven to system-driven control. Future research could explore system-driven interface adaptation based on performance metrics or user modeling, as well as mixed-initiative approaches where the system suggests level changes or adapts specific interface elements contextually. 
Second, expert feedback suggested that they desired UI customization for personalized workflows. This encourages the exploration of combining on-demand and scaffolded UI generation. Future research could use our technical pipeline, which allows interface iteration with human control, to let users customize their interfaces and evaluate such dynamic user experience.
Third, we evaluated task-conditioned scaffolded interfaces in Blender for 3D modeling and animation tasks. As introduced in Section~\ref{sec:generalizability} and discussed in Section~\ref{sec:discussion_generalizability}, our method can be used to generate scaffolded UIs for broader tasks and in other software tools. Future research could evaluate scaffolded interfaces in other applications and domains to explore broader use and impact.
Lastly, we tested GPT-4o as the LLM to analyze user task, select tools, identify domain concepts, and generate UI implementation code. Future work can evaluate our method using more advanced LLMs, which may reason tools and concepts that are more aligned with UI designers', software educators', or end users' requirements with improved reasoning and instruction following capabilities. In addition, LLMs adept at generating code can help to boost the UI implementation reliability, thus enabling easier UI iteration and customization.

\section{Conclusion}
Steep learning curves in professional creative software due to complex interfaces significantly hinder user learning and skill development. In this paper, we introduced \papername{}, a method for generating scaffolded interfaces in professional creative software to support task-based learning. The scaffolded interfaces present task-relevant tools, manage complexity through adjustable levels, organize the UI around workflow stages with domain concepts integrated, and connect users to native software interactions to support skill transfer. Our evaluation involving both beginners and experts showed that the scaffolded interfaces significantly improved task performance and efficiency for both user groups, supported concept understanding and learning during task execution for beginners, and demonstrated potential to create personalized UIs to augment productivity and creativity for experts. These findings showed the effectiveness of combining task-awareness, workflow-based organization, concept integration, and progressive feature disclosure in interface design. Based on a discussion of the study results, we provided insights for future research on user empowerment through scaffolded interfaces for software learners, educators, and end users.

\begin{acks}
We thank Jennifer Jacobs and Yuan-Fang Wang for their helpful suggestions on this project, our user study participants for their time and the valuable insights they contributed to our evaluation, and the anonymous reviewers whose constructive feedback helped us improve this paper.
\end{acks}

\bibliographystyle{ACM-Reference-Format}
\bibliography{bibliography}

@incollection{sweller2011cognitive,
  title={Cognitive load theory},
  author={Sweller, John},
  booktitle={Psychology of learning and motivation},
  volume={55},
  pages={37--76},
  year={2011},
  publisher={Elsevier}
}

@online{uiandvis1999hearst,
  author = {Marti Hearst},
  title = {Modern Information Retrieval---User Interfaces and Visualization},
  year = 1999,
  url = {https://people.ischool.berkeley.edu/~hearst/irbook/10/},
  update = {2025-04},
  organization={Addison-Wesley-Longman Publishing Co.}
}

@book{belland2017instructional,
  title={Instructional scaffolding in STEM education: Strategies and efficacy evidence},
  author={Belland, Brian R},
  year={2017},
  publisher={Springer Nature}
}

@online{krol2025copilotvision,
  author = {Jacob Krol},
  title = {Windows is about to get its biggest intelligent upgrade thanks to Copilot},
  year = 2025,
  url = {https://www.techradar.com/computing/artificial-intelligence/windows-is-about-to-get-its-biggest-intelligent-upgrade-thanks-to-copilot},
  update = {2025-04},
  organization={Tech Radar}
}

@online{cunningham2025copilotvision,
  author = {Andrew Cunningham},
  title = {Windows 11’s Copilot Vision wants to help you learn to use complicated apps},
  year = 2025,
  url = {https://arstechnica.com/gadgets/2025/04/windows-11s-copilot-vision-wants-to-help-you-learn-to-use-complicated-apps/},
  update = {2025-04},
  organization={ARS Technica}
}

@online{google2025aistudio,
  author = {Google},
  title = {Google AI Studio},
  year = 2025,
  url = {https://aistudio.google.com/},
  update = {2025-04},
  organization={Google}
}

@inproceedings{hart2006nasa,
  title={NASA-task load index (NASA-TLX); 20 years later},
  author={Hart, Sandra G},
  booktitle={Proceedings of the human factors and ergonomics society annual meeting},
  volume={50},
  number={9},
  pages={904--908},
  year={2006},
  organization={Sage publications Sage CA: Los Angeles, CA}
}

@book{keller2009motivational,
  title={Motivational design for learning and performance: The ARCS model approach},
  author={Keller, John M},
  year={2009},
  publisher={Springer Science \& Business Media}
}

@online{keller2010imms,
  author = {Keller, John M},
  title = {Instructional Materials Motivation Survey},
  year = 2025,
  url = {https://learninglab.uni-due.de/research-instrument/13887},
  update = {2025-04},
  organization={UNIVERSITÄT DUISBURG-ESSEN}
}

@article{kirakowski1996software,
  title={The software usability measurement inventory: background and usage},
  author={Kirakowski, Jurek},
  journal={Usability evaluation in industry},
  pages={169--178},
  year={1996}
}

@online{photoshop,
  author = {Adobe Photoshop},
  title = {Adobe Photoshop},
  year = 2025,
  url = {https://www.adobe.com/products/photoshop.html},
  update = {2025-4},
  organization={Adobe}
}

@online{illustrator,
  author = {Adobe Illustrator},
  title = {Adobe Illustrator},
  year = 2025,
  url = {https://www.adobe.com/products/illustrator.html},
  update = {2025-4},
  organization={Adobe}
}

@online{blender,
  author = {Blender},
  title = {Blender},
  year = 2025,
  url = {https://www.blender.org/},
  update = {2025-4},
  organization={Blender}
}

@online{maya,
  author = {Autodesk Maya},
  title = {Autodesk Maya},
  year = 2025,
  url = {https://www.autodesk.com/products/maya/},
  update = {2025-4},
  organization={Autodesk}
}

@online{tinkercad,
  author = {Autodesk},
  title = {Autodesk TinkerCAD},
  year = 2025,
  url = {https://www.tinkercad.com/},
  update = {2025-4},
  organization={Autodesk}
}

@online{illustratoripad,
  author = {Adobe},
  title = {Illustrator on the iPad. Draw on inspiration anywhere.},
  year = 2025,
  url = {https://www.adobe.com/products/illustrator/ipad.html?msockid=36b836cad4cf6d4502aa20c7d5086c30},
  update = {2025-9},
  organization={Adobe}
}

@inproceedings{chilana2018supporting,
  title={Supporting remote real-time expert help: Opportunities and challenges for novice 3d modelers},
  author={Chilana, Parmit K and Hudson, Nathaniel and Bhaduri, Srinjita and Shashikumar, Prashant and Kane, Shaun},
  booktitle={2018 IEEE Symposium on Visual Languages and Human-Centric Computing (VL/HCC)},
  pages={157--166},
  year={2018},
  organization={IEEE}
}

@article{lee2010usability,
  title={Usability principles and best practices for the user interface design of complex 3D architectural design and engineering tools},
  author={Lee, Ghang and Eastman, Charles M and Taunk, Tarang and Ho, Chun-Heng},
  journal={International journal of human-computer studies},
  volume={68},
  number={1-2},
  pages={90--104},
  year={2010},
  publisher={Elsevier}
}

@inproceedings{mcgrenere2002evaluation,
  title={An evaluation of a multiple interface design solution for bloated software},
  author={McGrenere, Joanna and Baecker, Ronald M and Booth, Kellogg S},
  booktitle={Proceedings of the SIGCHI conference on Human factors in computing systems},
  pages={164--170},
  year={2002}
}

@inproceedings{grossman2010chronicle,
  title={Chronicle: capture, exploration, and playback of document workflow histories},
  author={Grossman, Tovi and Matejka, Justin and Fitzmaurice, George},
  booktitle={Proceedings of the 23nd annual ACM symposium on User interface software and technology},
  pages={143--152},
  year={2010}
}

@article{braun2006using,
  title={Using thematic analysis in psychology},
  author={Braun, Virginia and Clarke, Victoria},
  journal={Qualitative research in psychology},
  volume={3},
  number={2},
  pages={77--101},
  year={2006},
  publisher={Taylor \& Francis}
}

@article{carroll1984training,
  title={Training wheels in a user interface},
  author={Carroll, John M and Carrithers, Caroline},
  journal={Communications of the ACM},
  volume={27},
  number={8},
  pages={800--806},
  year={1984},
  publisher={ACM New York, NY, USA}
}

@incollection{jameson2007adaptive,
  title={Adaptive interfaces and agents},
  author={Jameson, Anthony},
  booktitle={The human-computer interaction handbook},
  pages={459--484},
  year={2007},
  publisher={CRC press}
}

@inproceedings{chilana2012lemonaid,
  title={LemonAid: selection-based crowdsourced contextual help for web applications},
  author={Chilana, Parmit K and Ko, Amy J and Wobbrock, Jacob O},
  booktitle={Proceedings of the SIGCHI Conference on Human Factors in Computing Systems},
  pages={1549--1558},
  year={2012}
}

@inproceedings{khurana2024and,
  title={Why and when llm-based assistants can go wrong: Investigating the effectiveness of prompt-based interactions for software help-seeking},
  author={Khurana, Anjali and Subramonyam, Hariharan and Chilana, Parmit K},
  booktitle={Proceedings of the 29th International Conference on Intelligent User Interfaces},
  pages={288--303},
  year={2024}
}

@inproceedings{fraser2019replay,
  title={RePlay: contextually presenting learning videos across software applications},
  author={Fraser, C Ailie and Ngoon, Tricia J and Dontcheva, Mira and Klemmer, Scott},
  booktitle={Proceedings of the 2019 CHI Conference on Human Factors in Computing Systems},
  pages={1--13},
  year={2019}
}

@inproceedings{yang2024aqua,
  title={AQuA: Automated Question-Answering in Software Tutorial Videos with Visual Anchors},
  author={Yang, Saelyne and Vermeulen, Jo and Fitzmaurice, George and Matejka, Justin},
  booktitle={Proceedings of the 2024 CHI Conference on Human Factors in Computing Systems},
  pages={1--19},
  year={2024}
}

@inproceedings{zhong2021helpviz,
  title={Helpviz: Automatic generation of contextual visual mobile tutorials from text-based instructions},
  author={Zhong, Mingyuan and Li, Gang and Chi, Peggy and Li, Yang},
  booktitle={The 34th Annual ACM Symposium on User Interface Software and Technology},
  pages={1144--1153},
  year={2021}
}

@inproceedings{yang2022softvideo,
  title={SoftVideo: Improving the Learning Experience of Software Tutorial Videos with Collective Interaction Data},
  author={Yang, Saelyne and Yim, Jisu and Baigutanova, Aitolkyn and Kim, Seoyoung and Chang, Minsuk and Kim, Juho},
  booktitle={Proceedings of the 27th International Conference on Intelligent User Interfaces},
  pages={646--660},
  year={2022}
}

@inbook{khurana2025me,
author = {Khurana, Anjali and Su, Xiaotian and Wang, April Yi and Chilana, Parmit K},
title = {Do It For Me vs. Do It With Me: Investigating User Perceptions of Different Paradigms of Automation in Copilots for Feature-Rich Software},
year = {2025},
isbn = {9798400713941},
publisher = {Association for Computing Machinery},
address = {New York, NY, USA},
url = {https://doi.org/10.1145/3706598.3713431},
booktitle = {Proceedings of the 2025 CHI Conference on Human Factors in Computing Systems},
articleno = {880},
numpages = {18}
}

@article{horvitz2013lumiere,
  title={The Lumiere project: Bayesian user modeling for inferring the goals and needs of software users},
  author={Horvitz, Eric J and Breese, John S and Heckerman, David and Hovel, David and Rommelse, Koos},
  journal={arXiv preprint arXiv:1301.7385},
  year={2013}
}

@inproceedings{matejka2011ambient,
  title={Ambient help},
  author={Matejka, Justin and Grossman, Tovi and Fitzmaurice, George},
  booktitle={Proceedings of the SIGCHI Conference on Human Factors in Computing Systems},
  pages={2751--2760},
  year={2011}
}

@online{firefly,
  author = {Adobe},
  title = {Adobe Firefly},
  year = 2025,
  url = {https://www.adobe.com/products/firefly.html},
  update = {2025-4},
  organization={Adobe}
}

@misc{liu2025crowdgenui,
  title={CrowdGenUI: Aligning LLM-Based UI Generation with Crowdsourced User Preferences}, 
  author={Yimeng Liu and Misha Sra and Chang Xiao},
  year={2025},
  eprint={2411.03477},
  archivePrefix={arXiv},
  primaryClass={cs.HC},
  url={https://arxiv.org/abs/2411.03477}, 
}

@inproceedings{vaithilingam2024dynavis,
  title={Dynavis: Dynamically synthesized ui widgets for visualization editing},
  author={Vaithilingam, Priyan and Glassman, Elena L and Inala, Jeevana Priya and Wang, Chenglong},
  booktitle={Proceedings of the 2024 CHI Conference on Human Factors in Computing Systems},
  pages={1--17},
  year={2024}
}

@inproceedings{li2024user,
  title={User experience design professionals’ perceptions of generative artificial intelligence},
  author={Li, Jie and Cao, Hancheng and Lin, Laura and Hou, Youyang and Zhu, Ruihao and El Ali, Abdallah},
  booktitle={Proceedings of the 2024 CHI Conference on Human Factors in Computing Systems},
  pages={1--18},
  year={2024}
}

@inproceedings{kiani2019beyond,
  title={Beyond" One-Size-Fits-All" Understanding the Diversity in How Software Newcomers Discover and Make Use of Help Resources},
  author={Kiani, Kimia and Cui, George and Bunt, Andrea and McGrenere, Joanna and Chilana, Parmit K},
  booktitle={Proceedings of the 2019 CHI Conference on Human Factors in Computing Systems},
  pages={1--14},
  year={2019}
}

@article{endsley2017here,
  title={From here to autonomy: lessons learned from human--automation research},
  author={Endsley, Mica R},
  journal={Human factors},
  volume={59},
  number={1},
  pages={5--27},
  year={2017},
  publisher={Sage Publications Sage CA: Los Angeles, CA}
}

@article{heer2019agency,
  title={Agency plus automation: Designing artificial intelligence into interactive systems},
  author={Heer, Jeffrey},
  journal={Proceedings of the National Academy of Sciences},
  volume={116},
  number={6},
  pages={1844--1850},
  year={2019},
  publisher={National Academy of Sciences}
}

@article{sellen2024rise,
  title={The rise of the ai co-pilot: Lessons for design from aviation and beyond},
  author={Sellen, Abigail and Horvitz, Eric},
  journal={Communications of the ACM},
  volume={67},
  number={7},
  pages={18--23},
  year={2024},
  publisher={ACM New York, NY, USA}
}

@article{rieman1996field,
  title={A field study of exploratory learning strategies},
  author={Rieman, John},
  journal={ACM Transactions on Computer-Human Interaction (TOCHI)},
  volume={3},
  number={3},
  pages={189--218},
  year={1996},
  publisher={ACM New York, NY, USA}
}

@inproceedings{novick2009micro,
  title={The micro-structure of use of help},
  author={Novick, David G and Andrade, Oscar D and Bean, Nathaniel},
  booktitle={Proceedings of the 27th ACM international conference on Design of communication},
  pages={97--104},
  year={2009}
}

@article{farkas1993role,
  title={The role of balloon help},
  author={Farkas, David K},
  journal={ACM SIGDOC Asterisk Journal of Computer Documentation},
  volume={17},
  number={2},
  pages={3--19},
  year={1993},
  publisher={ACM New York, NY, USA}
}

@inproceedings{grossman2010toolclips,
  title={ToolClips: an investigation of contextual video assistance for functionality understanding},
  author={Grossman, Tovi and Fitzmaurice, George},
  booktitle={Proceedings of the SIGCHI Conference on Human Factors in Computing Systems},
  pages={1515--1524},
  year={2010}
}

@inproceedings{fourney2014intertwine,
  title={InterTwine: creating interapplication information scent to support coordinated use of software},
  author={Fourney, Adam and Lafreniere, Ben and Chilana, Parmit and Terry, Michael},
  booktitle={Proceedings of the 27th annual ACM symposium on User interface software and technology},
  pages={429--438},
  year={2014}
}

@inproceedings{kelleher2005stencils,
  title={Stencils-based tutorials: design and evaluation},
  author={Kelleher, Caitlin and Pausch, Randy},
  booktitle={Proceedings of the SIGCHI conference on Human factors in computing systems},
  pages={541--550},
  year={2005}
}

@inproceedings{matejka2011ip,
  title={IP-QAT: in-product questions, answers, \& tips},
  author={Matejka, Justin and Grossman, Tovi and Fitzmaurice, George},
  booktitle={Proceedings of the 24th annual ACM symposium on User interface software and technology},
  pages={175--184},
  year={2011}
}

@article{li2005active,
  title={Active affective state detection and user assistance with dynamic Bayesian networks},
  author={Li, Xiangyang and Ji, Qiang},
  journal={IEEE transactions on systems, man, and cybernetics-part a: systems and humans},
  volume={35},
  number={1},
  pages={93--105},
  year={2005},
  publisher={IEEE}
}

@inproceedings{ekstrand2011searching,
  title={Searching for software learning resources using application context},
  author={Ekstrand, Michael and Li, Wei and Grossman, Tovi and Matejka, Justin and Fitzmaurice, George},
  booktitle={Proceedings of the 24th annual ACM symposium on User interface software and technology},
  pages={195--204},
  year={2011}
}

@inproceedings{novick2006don,
  title={Why don't people read the manual?},
  author={Novick, David G and Ward, Karen},
  booktitle={Proceedings of the 24th annual ACM international conference on Design of communication},
  pages={11--18},
  year={2006}
}

@inproceedings{lafreniere2013community,
  title={Community enhanced tutorials: improving tutorials with multiple demonstrations},
  author={Lafreniere, Benjamin and Grossman, Tovi and Fitzmaurice, George},
  booktitle={Proceedings of the SIGCHI Conference on Human Factors in Computing Systems},
  pages={1779--1788},
  year={2013}
}

@article{cockburn2014supporting,
author = {Cockburn, Andy and Gutwin, Carl and Scarr, Joey and Malacria, Sylvain},
year = {2014},
month = {01},
pages = {1-36},
title = {Supporting Novice to Expert Transitions in User Interfaces},
volume = {47},
journal = {ACM Computing Surveys},
doi = {10.1145/2659796}
}

@online{nicol2024psychology,
  author = {Shaan Nicol},
  title = {The Psychology Behind Effective UI/UX Design},
  year = 2024,
  url = {https://www.chillybin.co/psychology-ui-ux-design/},
  update = {2024-10},
  organization={ChillyBin}
}

@article{li2022design,
  title={Design factors to improve the consistency and sustainable user experience of responsive Interface Design},
  author={Li, Wenjie and Zhou, Yuxiao and Luo, Shijian and Dong, Yenan},
  journal={Sustainability},
  volume={14},
  number={15},
  pages={9131},
  year={2022},
  publisher={MDPI}
}

@article{brdnik2022intelligent,
  title={Intelligent user interfaces and their evaluation: a systematic mapping study},
  author={Brdnik, Sa{\v{s}}a and Heri{\v{c}}ko, Tja{\v{s}}a and {\v{S}}umak, Bo{\v{s}}tjan},
  journal={Sensors},
  volume={22},
  number={15},
  pages={5830},
  year={2022},
  publisher={MDPI}
}

@article{kosch2023survey,
  title={A survey on measuring cognitive workload in human-computer interaction},
  author={Kosch, Thomas and Karolus, Jakob and Zagermann, Johannes and Reiterer, Harald and Schmidt, Albrecht and Wo{\'z}niak, Pawe{\l} W},
  journal={ACM Computing Surveys},
  volume={55},
  number={13s},
  pages={1--39},
  year={2023},
  publisher={ACM New York, NY}
}

@book{lim2012improving,
  title={Improving understanding and trust with intelligibility in context-aware applications},
  author={Lim, Brian Y},
  year={2012},
  publisher={Carnegie Mellon University}
}

@techreport{vorvoreanu2025fostering,
author = {Vorvoreanu, Mihaela and Passi, Samir and Dhanorkar, Shipi and Heger, Amy and Walker, Kathleen},
title = {Fostering appropriate reliance on GenAI: Lessons learned from early research},
institution = {Microsoft},
year = {2025},
month = {March},
url = {https://www.microsoft.com/en-us/research/publication/fostering-appropriate-reliance-on-genai-lessons-learned-from-early-research/},
number = {MSR-TR-2025-4},
}

@article{chen2023understanding,
  title={Understanding the role of human intuition on reliance in human-AI decision-making with explanations},
  author={Chen, Valerie and Liao, Q Vera and Wortman Vaughan, Jennifer and Bansal, Gagan},
  journal={Proceedings of the ACM on Human-computer Interaction},
  volume={7},
  number={CSCW2},
  pages={1--32},
  year={2023},
  publisher={ACM New York, NY, USA}
}

@inproceedings{lafreniere2014task,
  title={Task-centric interfaces for feature-rich software},
  author={Lafreniere, Benjamin and Bunt, Andrea and Terry, Michael},
  booktitle={Proceedings of the 26th Australian Computer-Human Interaction Conference on Designing Futures: The Future of Design},
  pages={49--58},
  year={2014}
}

@article{qiao2024benchmarking,
  title={Benchmarking Agentic Workflow Generation},
  author={Qiao, Shuofei and Fang, Runnan and Qiu, Zhisong and Wang, Xiaobin and Zhang, Ningyu and Jiang, Yong and Xie, Pengjun and Huang, Fei and Chen, Huajun},
  journal={arXiv preprint arXiv:2410.07869},
  year={2024}
}

@inproceedings{seabold2010statsmodels,
  title={statsmodels: Econometric and statistical modeling with python},
  author={Seabold, Skipper and Perktold, Josef},
  booktitle={Proceedings of the 9th Python in Science Conference},
  pages={57--61},
  year={2010}
}

@article{Vallat2018pingouin,
  doi = {10.21105/joss.01026},
  url = {https://doi.org/10.21105/joss.01026},
  year = {2018},
  publisher = {The Open Journal},
  volume = {3},
  number = {31},
  pages = {1026},
  author = {Raphael Vallat},
  title = {Pingouin: statistics in Python},
  journal = {Journal of Open Source Software}
}

@inproceedings{liu2024dancegen,
  title={DanceGen: Supporting Choreography Ideation and Prototyping with Generative AI},
  author={Liu, Yimeng and Sra, Misha},
  booktitle={Proceedings of the 2024 ACM Designing Interactive Systems Conference},
  pages={920--938},
  year={2024}
}

@article{poole2022dreamfusion,
  author = {Poole, Ben and Jain, Ajay and Barron, Jonathan T. and Mildenhall, Ben},
  title = {DreamFusion: Text-to-3D using 2D Diffusion},
  journal = {arXiv},
  year = {2022},
}

@inproceedings{qin2024instructvid2vid,
  title={Instructvid2vid: Controllable video editing with natural language instructions},
  author={Qin, Bosheng and Li, Juncheng and Tang, Siliang and Chua, Tat-Seng and Zhuang, Yueting},
  booktitle={2024 IEEE International Conference on Multimedia and Expo (ICME)},
  pages={1--6},
  year={2024},
  organization={IEEE}
}

@online{generativefill,
  author = {Adobe},
  title = {Adobe Photoshop Generative Fill},
  year = 2025,
  url = {https://www.adobe.com/products/photoshop/generative-fill.html},
  update = {2025-5},
  organization={Adobe}
}

@online{knowledge_navigator,
  author = {Apple},
  title = {Knowledge Navigator},
  year = 1987,
  url = {https://en.wikipedia.org/wiki/Knowledge_Navigator},
  update = {2025-5},
  organization={Apple Inc.}
}

@article{shneiderman2002promoting,
  title={Promoting universal usability with multi-layer interface design},
  author={Shneiderman, Ben},
  journal={ACM SIGCAPH computers and the physically handicapped},
  number={73-74},
  pages={1--8},
  year={2002},
  publisher={ACM New York, NY, USA}
}

@article{kosmyna2025your,
  title={Your Brain on ChatGPT: Accumulation of Cognitive Debt when Using an AI Assistant for Essay Writing Task},
  author={Kosmyna, Nataliya and Hauptmann, Eugene and Yuan, Ye Tong and Situ, Jessica and Liao, Xian-Hao and Beresnitzky, Ashly Vivian and Braunstein, Iris and Maes, Pattie},
  journal={arXiv preprint arXiv:2506.08872},
  year={2025}
}

@online{blendermanuals,
  author = {Blender},
  title = {Blender 3.6 Reference Manual},
  year = 2025,
  url = {https://docs.blender.org/manual/en/3.6/index.html#},
  update = {2025-6},
  organization={Blender}
}

@online{blenderapidocs,
  author = {Blender},
  title = {Blender 3.6 Python API Documentation},
  year = 2025,
  url = {https://docs.blender.org/api/3.6/},
  update = {2025-6},
  organization={Blender}
}

@inproceedings{popovic2000expert,
  title={Expert and novice user differences and implications for product design and useability},
  author={Popovic, Vesna},
  booktitle={Proceedings of the Human Factors and Ergonomics Society Annual Meeting},
  volume={44},
  number={38},
  pages={933--936},
  year={2000},
  organization={SAGE Publications Sage CA: Los Angeles, CA}
}

@article{ruginski2019gps,
  title={GPS use negatively affects environmental learning through spatial transformation abilities},
  author={Ruginski, Ian T and Creem-Regehr, Sarah H and Stefanucci, Jeanine K and Cashdan, Elizabeth},
  journal={Journal of Environmental Psychology},
  volume={64},
  pages={12--20},
  year={2019},
  publisher={Elsevier}
}

@online{blender_copilot,
  author = {Blender},
  title = {Blender Copilot},
  year = 2025,
  url = {https://blenderside.gumroad.com/l/blender-copilot},
  update = {2025-11},
  organization={Blender}
}

@online{blender_copilot_oai,
  author = {OpenAI},
  title = {Blender Copilot 5.1},
  year = 2025,
  url = {https://chatgpt.com/g/g-w7Ewzzi94-blender-copilot},
  update = {2025-11},
  organization={OpenAI}
}

@inproceedings{yuan2025understanding,
  title={Understanding and mitigating numerical sources of nondeterminism in llm inference},
  author={Yuan, Jiayi and Li, Hao and Ding, Xinheng and Xie, Wenya and Li, Yu-Jhe and Zhao, Wentian and Wan, Kun and Shi, Jing and Hu, Xia and Liu, Zirui},
  booktitle={The Thirty-ninth Annual Conference on Neural Information Processing Systems},
  year={2025}
}

@article{zhou2023instruction,
  title={Instruction-following evaluation for large language models},
  author={Zhou, Jeffrey and Lu, Tianjian and Mishra, Swaroop and Brahma, Siddhartha and Basu, Sujoy and Luan, Yi and Zhou, Denny and Hou, Le},
  journal={arXiv preprint arXiv:2311.07911},
  year={2023}
}

@online{swanson2026aifluency,
author = {Kristen Swanson and Drew Bent and Zoe Ludwig and Rick Dakan and Joe Feller},
title = {Anthropic Education Report: The AI Fluency Index},
date = {2026-02-16},
year = {2026},
url = {https://www.anthropic.com/news/anthropic-education-report-the-ai-fluency-index}
}

@article{leviathangenerative,
  title={Generative UI: LLMs are Effective UI Generators},
  author={Leviathan, Yaniv and Kalman, Dani Valevski Matan and Lumen, Danny and Molad, Eyal Segalis Eyal and Pasternak, Shlomi and Natchu, Vishnu and Nygaard, Valerie and Matias, Srinivasan Cheenu Venkatachary James Manyika Yossi}
}

@article{cheng2024biscuit,
  title={BISCUIT: Scaffolding LLM-Generated Code with Ephemeral UIs in Computational Notebooks},
  author={Cheng, Ruijia and Barik, Titus and Leung, Alan and Hohman, Fred and Nichols, Jeffrey},
  journal={arXiv preprint arXiv:2404.07387},
  year={2024}
}

@inproceedings{cao2025generative,
  title={Generative and malleable user interfaces with generative and evolving task-driven data model},
  author={Cao, Yining and Jiang, Peiling and Xia, Haijun},
  booktitle={Proceedings of the 2025 CHI Conference on Human Factors in Computing Systems},
  pages={1--20},
  year={2025}
}

@inproceedings{min2025malleable,
  title={Malleable overview-detail interfaces},
  author={Min, Bryan and Chen, Allen and Cao, Yining and Xia, Haijun},
  booktitle={Proceedings of the 2025 CHI Conference on Human Factors in Computing Systems},
  pages={1--25},
  year={2025}
}

@inproceedings{xu2025productive,
  title={Productive vs. reflective: how different ways of integrating AI into design workflows affect cognition and motivation},
  author={Xu, Xiaotong and Konnova, Arina and Gao, Bianca and Peng, Cindy and Vo, Dave and Dow, Steven P},
  booktitle={Proceedings of the 2025 CHI Conference on Human Factors in Computing Systems},
  pages={1--15},
  year={2025}
}

@inproceedings{bergman2005docwizards,
  title={DocWizards: a system for authoring follow-me documentation wizards},
  author={Bergman, Lawrence and Castelli, Vittorio and Lau, Tessa and Oblinger, Daniel},
  booktitle={Proceedings of the 18th annual ACM symposium on User interface software and technology},
  pages={191--200},
  year={2005}
}

@inproceedings{huang2007graphstract,
  title={Graphstract: minimal graphical help for computers},
  author={Huang, Jeff and Twidale, Michael B},
  booktitle={Proceedings of the 20th annual ACM symposium on User interface software and technology},
  pages={203--212},
  year={2007}
}

\appendix
\onecolumn

\section{LLM Prompt} \label{sec:llm_prompt}
\subsection{Workflow Analysis and Concept Identification} \label{sec:llm_prompt_workflow}

\begin{llmscriptblock}{}
\noindent You are an expert task workflow analyst specializing in task decomposition for a user task. Your objective is to analyze the user task and break it down into its primary workflow stages.
\vspace{1em}

\noindent\textbf{TASK TO ANALYZE:}
\par\noindent
\llmplaceholder{USER TASK DESCRIPTION}
\vspace{1em}

\noindent\textbf{METHOD:}
\par\noindent
Decompose the task into logical stages that represent a typical, efficient workflow a user would follow within \llmplaceholder{SOFTWARE}. Focus on the major distinct phases, moving from initial setup/preparation towards finalization/review. Aim for a reasonable number of stages that capture the core workflow without excessive granularity. Also, extract domain concepts relevant to each workflow stage. For each domain concepts, provide a concise, straightforward explanation for a human to easily understand. 
\vspace{1em}

\noindent\textbf{OUTPUT REQUIREMENTS:}
\par\noindent
Provide the output as a numbered list. Each list item must represent a distinct workflow stage and include:
A concise \textbf{Stage Name}.
A brief, one-sentence \textbf{Stage Description} clarifying the purpose of the stage.
\textbf{Domain Concepts} related to the stage. 
A concise, straightforward \textbf{Concept Explanation} explaining the concept. 
\vspace{0.5em}

\noindent
The output must be structured following the EXAMPLE OUTPUT FORMAT.
\vspace{1em}

\noindent\textbf{EXAMPLE OUTPUT FORMAT:}
\par
\begin{itemize}[label=\textendash, wide, nosep, leftmargin=1.5em, topsep=2pt]
    \item \textbf{Stage Name 1}: Stage Description, Domain Concepts, Concept Explanation.
    \item \textbf{Stage Name 2}: Stage Description, Domain Concepts, Concept Explanation.
    \item \dots
    \item \textbf{Stage Name n}: Stage Description, Domain Concepts, Concept Explanation.
\end{itemize}
\end{llmscriptblock}

\subsection{Tool Selection and Complexity Assessment} \label{sec:llm_prompt_tool}

\begin{llmscriptblock}{}
\noindent You are an expert user of \llmplaceholder{SOFTWARE}. Your task is to identify relevant tools in \llmplaceholder{SOFTWARE} to perform each workflow stage of \llmplaceholder{WORKFLOW STAGES}.
\vspace{1em}

\noindent\textbf{WORKFLOW STAGES:}
\begin{itemize}[label=\textendash, wide, nosep, leftmargin=2em]
    \item \llmplaceholder{STAGE NAME}: \llmplaceholder{STAGE DESCRIPTION}, \llmplaceholder{DOMAIN CONCEPTS}, \llmplaceholder{CONCEPT EXPLANATION}
\end{itemize}
\vspace{1em}

\noindent\textbf{METHOD:}
\par\noindent
Identify the keyboard operators, mouse operations, and UI interactions needed to perform each workflow stage. The identified tools must be common and essential for the workflow stage. Prioritize stable and frequently used tools. In addition, you need to assign a difficulty level for each identified tool by referring to \llmplaceholder{SOFTWARE MANUALS}. The difficulty level should be basic, intermediate, or advanced for a human user. 
\vspace{1em}

\noindent\textbf{OUTPUT REQUIREMENTS:}
\par\noindent
For each stage, output a bulleted list of selected tools. Each item in the list must include: 
The \textbf{Stage Name} from the \llmplaceholder{WORKFLOW STAGES}. 
The \textbf{Selected Tools} required to perform the task in the stage. 
A \textbf{Rationale} that briefly and clearly explain why the selected tools are essential and necessary. 
A \textbf{Complexity Level} of Basic, Intermediate, or Advanced for the selected tools. 
\vspace{1em}

\noindent\textbf{EXAMPLE OUTPUT FORMAT:}
\par
\begin{itemize}[label=\textendash, wide, nosep, leftmargin=1.5em, topsep=2pt]
    \item \textbf{Stage Name 1}: Selected Tools, Rationale, Complexity Level.
    \item \textbf{Stage Name 2}: Selected Tools, Rationale, Complexity Level.
    \item \dots
    \item \textbf{Stage Name n}: Selected Tools, Rationale, Complexity Level.
\end{itemize}
\end{llmscriptblock}

\subsection{UI Code Generation and Tool Labeling} \label{sec:llm_prompt_code}

\begin{llmscriptblock}{}
\noindent You are a coding expert for \llmplaceholder{SOFTWARE}. Your task is to generate the code to implement \llmplaceholder{SELECTED TOOLS} based on \llmplaceholder{SOFTWARE} functionality and generate UI code to display \llmplaceholder{SELECTED TOOLS} in a user interface for user interaction.
\vspace{1em}

\noindent\textbf{SELECTED TOOLS:}
\begin{itemize}[label=\textendash, wide, nosep, leftmargin=1.5em, topsep=2pt]
    \item \llmplaceholder{STAGE NAME}: \llmplaceholder{SELECTED TOOLS}, \llmplaceholder{RATIONALE}, \llmplaceholder{COMPLEXITY LEVEL}.
\end{itemize}
\vspace{1em}

\noindent\textbf{METHOD:}
\par\noindent
For each selected tool in \llmplaceholder{SELECTED TOOLS}, you need to translate the tool operation into the corresponding code execution by referring to \llmplaceholder{SOFTWARE APIS}. You must make sure the code can achieve the identical result using the tool. The code you generate should include all the required parameters and properties for the code to be correctly executed. 
\vspace{0.5em}

\noindent
Next, you need to convert the code into interactive UI controls that can be used by a human. You can use buttons, dropdown menus, radio buttons, tabs, text field, switch toggles, or other UI controls that are supported by \llmplaceholder{SOFTWARE} and suitable for each tool. You must refer to \llmplaceholder{SOFTWARE APIS} to make sure the UI control can be implemented. 
\vspace{0.5em}

\noindent
To organize the code for UI controls, you should refer to the \llmplaceholder{EXAMPLE CODE} to group the code for each UI control under their \llmplaceholder{WORKFLOW STAGE} and \llmplaceholder{COMPLEXITY LEVEL}. In addition, you should implement a mouse-over tooltip for each UI control with their relevant \llmplaceholder{DOMAIN CONCEPT} and \llmplaceholder{CONCEPT DESCRIPTION}. 
\vspace{1em}

\noindent\textbf{EXAMPLE CODE WITH UI LAYOUT STRUCTURE}
\par\noindent
\llmplaceholder{EXAMPLE CODE}
\vspace{1em}

\noindent\textbf{OUTPUT REQUIREMENTS:}
\par\noindent
Your output is a code script. The script should be standalone with all necessary imports, function definitions, and main function for it to be executable. You should refer to \llmplaceholder{EXAMPLE CODE} for the code script generation. 
\end{llmscriptblock}

\section{User Study Materials}

\subsection{Walk Cycle Scaffolded UI} \label{sec:use_scenario_s2_scaffold_ui}
\begin{figure*}[!ht]
    \centering
    \includegraphics[width=.9\textwidth]{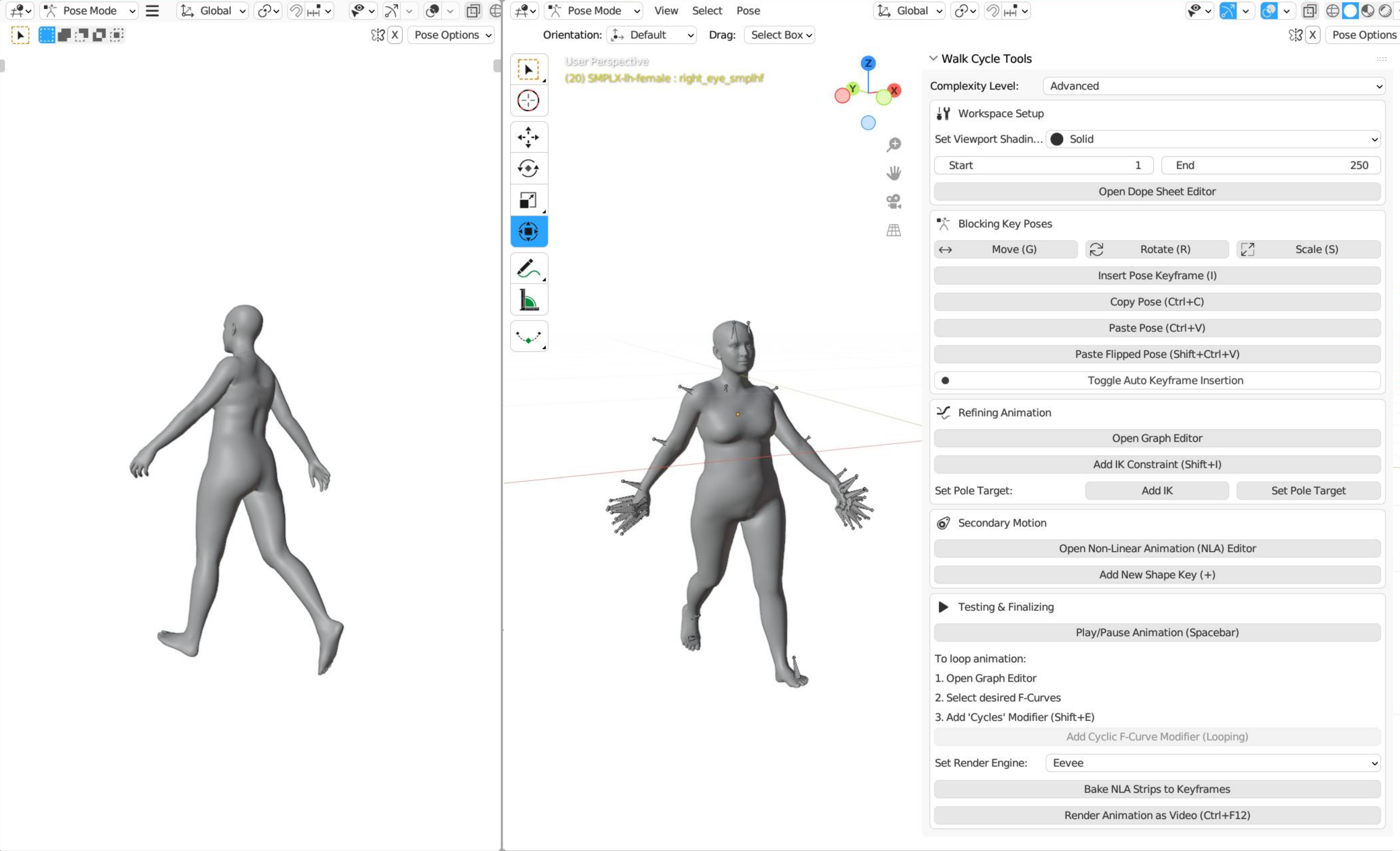}
    \caption{Our scaffolded UI generated to perform walk cycle animation.}
    \Description{This figure shows the scaffolded UI for building the walk cycle animation. This scaffolded UI appears as an add-on panel on the right side in Blender's UI.}
    \label{fig:use_scenario_s2_scaffold_ui}
\end{figure*}

Figure~\ref{fig:use_scenario_s2_scaffold_ui} shows the scaffolded UI for the walk cycle task. 

\subsection{Study Guide} \label{sec:study_guide}
The video tutorial for UV unwrapping is available at \url{https://www.youtube.com/watch?v=7JUNlj6mR0U}, and for building a walk cycle can be found at \url{https://www.youtube.com/watch?v=e_COc0ZVHr0}.

\subsection{Questionnaires} \label{sec:study1_questionnaires}
Table~\ref{tab:cognitive_load_questionnaire} summarizes the questions from NASA-TLX to evaluate task load. The custom questionnaires for task performance, concept learning, and user interface experience are presented in Tables~\ref{tab:task_performance_questionnaire},~\ref{tab:concept_learning_questionnaire}, and~\ref{tab:interface_experience_questionnaire}. 

\begin{table}[!ht]
    \centering
    \caption{Task load questionnaire (NASA-TLX).}
    \scalebox{0.95} {
    \begin{tabular}{p{0.27\textwidth}p{0.5\textwidth}}
    \toprule
        Mental demand & How mentally demanding was it using the interface to perform your task? \\ 
        \midrule
        Temporal demand & How hurried or rushed was it using the interface to perform your task? \\ 
        \midrule
        Performance & How successful were you in accomplishing what you were asked to do? \\ 
        \midrule
        Effort & How hard did you have to work to accomplish your level of performance? \\ 
        \midrule
        Frustration & How insecure, discouraged, irritated, stressed, and annoyed were you? \\
    \bottomrule
    \end{tabular}
    }
    \label{tab:cognitive_load_questionnaire}
\end{table}

\begin{table}[!ht]
    \centering
    \caption{Task performance questionnaire.}
    \scalebox{0.95} {
    \begin{tabular}{p{0.27\textwidth}p{0.5\textwidth}}
    \toprule
        Workflow clarity & I felt clear about the task workflow. \\ 
        \midrule
        Task efficiency & I was able to complete the task efficiently. \\ 
        \midrule
        Tool relevance & It was easy for me to find the tools I needed to perform my task. \\ 
        \midrule
        User focus \& engagement & It was easy for me to stay focused on and engaged in my task. \\ 
        \midrule
        User confidence \& satisfaction & I felt confident performing the task and was satisfied with my accomplishment. \\
    \bottomrule
    \end{tabular}
    }
    \label{tab:task_performance_questionnaire}
\end{table}

\begin{table}[!ht]
    \centering
    \caption{Concept learning questionnaire.}
    \scalebox{0.95} {
    \begin{tabular}{p{0.27\textwidth}p{0.5\textwidth}}
    \toprule
        Concept-workflow correlation & I was able to correlate task-related concepts with the task workflow. \\ 
        \midrule
        Concept understanding & Using this interface helped me to understand these task-related concepts. \\ 
        \midrule
        Concept-tool correlation & I was able to correlate tools used for my task with these task-related concepts. \\ 
        \midrule
        Learning motivation & Using this interface enhanced my motivation and experience of learning concepts. \\ 
        \midrule
        Adjacent concept exploration & I was able to explore concepts beyond those directly related to my task. \\
    \bottomrule
    \end{tabular}
    }
    \label{tab:concept_learning_questionnaire}
\end{table}

\begin{table}[!ht]
    \centering
    \caption{User interface experience questionnaire.}
    \scalebox{0.95} {
    \begin{tabular}{p{0.27\textwidth}p{0.5\textwidth}}
    \toprule
        Tool support vs. distraction & The interface offered relevant tools that supported my task without unnecessary distractions. \\ 
        \midrule
        Prior knowledge requirement & Prior interface familiarity (e.g., tool locations, usage, keyboard shortcuts) was not required to perform my task. \\ 
        \midrule
        External help requirement & Extra assistance (e.g., referring to tutorials or asking questions) was not required to perform my task. \\ 
        \midrule
        Adjacent tool exploration & I was able to explore tools in the interface beyond those directly related to my task. \\ 
        \midrule
        Interface vs. user adaptation & The interface appeared tailored to support my task rather than requiring me to adapt to it. \\
    \bottomrule
    \end{tabular}
    }
    \label{tab:interface_experience_questionnaire}
\end{table}

\subsection{Open-Ended Questions} \label{sec:study2_openended_questions}
\textit{Compared to the default Blender interface, please provide your experience with the custom interface. We would appreciate detailed responses.}

1. Did the custom interface help you complete the task better or worse than you expected? Were you able to easily find what you needed in the interface to perform your task? How did it make you feel while working, e.g., did you feel engaged or sometimes lost? Did the design of the interface make sense to you?

2. (Task 1: UV Unwrapping) If you were to learn new concepts, would the custom interface help you understand the concepts better or make them more confusing? For example, would things like seams, unwrapping, and UV stretching be easier to grasp? Would it help you discover anything new that isn't directly part of your task but a related idea such as UV islands? Would the way the interface presents the individual tools make learning feel natural or difficult?

2. (Task 2: Building Walk Cycle) If you were to learn new concepts, would the custom interface help you understand the concepts better or make them more confusing? For example, would things like key poses and bone constraints be easier to grasp? Would it help you discover anything new that isn't directly part of your task but a related idea such as secondary motion? Would the way the interface presents the individual tools make learning feel natural or difficult?

3. How did the custom interface affect your overall experience? Did it feel like it was supporting you or getting in your way? Could you figure things out on your own, or did you need help? Did you feel like you had to adjust to the system, or did the system adjust to you and your needs?

4. If you were to use the custom interface for a longer period, would it make your long-term skill development easier or harder? For example, would building up your skills from basic to advanced be easier, and if so, how? Would it help you build practical skills by teaching you underlying concepts? Would the learned concepts be transferable?

5. If you have experience with other software similar to Blender (e.g., Maya, 3ds Max), what do you think about the impact of using similar custom interfaces in them? How might the custom interfaces be used?

6. Are there any features you would add, remove, or modify in the custom interface? If yes, please specify.

\section{User Study 1 ANOVA Results} \label{sec:study1_anova}
Tables~\ref{tab:anova_taskload} to~\ref{tab:anova_interfaceexperience} report the two-way ANOVA results of Study 1.

\begin{table}[!ht]
 \centering
 \caption{Two-way ANOVA results for task load measures.}
 \label{tab:anova_taskload}
 \scalebox{0.95} {
 \begin{tabular}{p{0.27\textwidth} ccc ccc ccc}
 \toprule
 & \multicolumn{3}{c}{\textbf{Interface Effect}} & \multicolumn{3}{c}{\textbf{Task Effect}} & \multicolumn{3}{c}{\textbf{Interaction Effect}} \\
 \cmidrule(lr){2-4} \cmidrule(lr){5-7} \cmidrule(lr){8-10}
& $F_{(1, 30)}$ & $p$ & $\eta^{2}_{p}$& $F_{(1, 30)}$ & $p$ & $\eta^{2}_{p}$& $F_{(1, 30)}$ & $p$ & $\eta^{2}_{p}$\\
 \midrule
 Mental demand& $22.53$& $ < .001 $ & $.43$& $8.53$& $.007$& $.22$& $0.13$& $.718$& $.00$ \\
 Temporal demand & $95.38$& $ < .001 $ & $.76$& $4.71$& $.039$& $.14$& $0.52$& $.475$& $.02$ \\
 Performance& $150.36$ & $ < .001 $ & $.83$& $0.19$& $.665$& $.01$& $1.73$& $.200$& $.05$ \\
 Effort & $69.96$& $ < .001 $ & $.70$& $8.51$& $.007$& $.22$& $8.51$& $.007$& $.22$ \\
 Frustration& $91.43$& $ < .001 $ & $.75$& $4.32$& $.047$& $.13$& $2.77$& $.107$& $.08$ \\
 \bottomrule
 \end{tabular}
 }
\end{table}

\begin{table}[!ht]
 \centering
 \caption{Two-way ANOVA results for task performance measures.}
 \label{tab:anova_taskperformance}
\scalebox{0.95}{
\begin{tabular}{p{0.27\textwidth}ccc ccc ccc}
 \toprule
 & \multicolumn{3}{c}{\textbf{Interface Effect}} & \multicolumn{3}{c}{\textbf{Task Effect}} & \multicolumn{3}{c}{\textbf{Interaction Effect}} \\
 \cmidrule(lr){2-4} \cmidrule(lr){5-7} \cmidrule(lr){8-10}
 & $F_{(1, 30)}$ & $p$ & $\eta^{2}_{p}$& $F_{(1, 30)}$ & $p$ & $\eta^{2}_{p}$ & $F_{(1, 30)}$ & $p$ & $\eta^{2}_{p}$ \\
 \midrule
 Workflow clarity & $34.60$ & $ < .001 $ & $.54$ & $2.05$ & $.164$ & $.06$ & $0.63$ & $.433$ & $.02$ \\
 Task efficiency& $19.96$ & $ < .001 $ & $.40$ & $12.55$ & $.001$ & $.30$ & $0.21$ & $.648$ & $.01$ \\
 Tool relevance & $48.13$ & $ < .001 $ & $.62$ & $0.53$ & $.471$ & $.02$ & $0.13$ & $.718$ & $.00$ \\
 User focus \& engagement & $71.46$ & $ < .001 $ & $.70$ & $0.42$ & $.521$ & $.01$ & $2.30$ & $.140$ & $.07$ \\
 User confidence \& satisfaction & $38.53$ & $ < .001 $ & $.56$ & $2.13$ & $.155$ & $.07$ & $4.80$ & $.037$ & $.14$ \\
 \bottomrule
 \end{tabular}
 }
\end{table}

\begin{table}[!ht]
 \centering
 \caption{Two-way ANOVA results for concept learning measures.}
 \label{tab:anova_conceptlearning}
 \scalebox{0.95} {
 \begin{tabular}{p{0.27\textwidth} ccc ccc ccc}
 \toprule
 & \multicolumn{3}{c}{\textbf{Interface Effect}} & \multicolumn{3}{c}{\textbf{Task Effect}} & \multicolumn{3}{c}{\textbf{Interaction Effect}} \\
 \cmidrule(lr){2-4} \cmidrule(lr){5-7} \cmidrule(lr){8-10}
 & $F_{(1, 30)}$ & $p$ & $\eta^{2}_{p}$& $F_{(1, 30)}$ & $p$ & $\eta^{2}_{p}$ & $F_{(1, 30)}$ & $p$ & $\eta^{2}_{p}$ \\
 \midrule
 Concept-workflow correlation & $78.91$ & $ < .001 $ & $.73$ & $7.39$ & $.011$ & $.20$ & $0.03$ & $.857$ & $.00$ \\
 Concept understanding& $29.67$ & $ < .001 $ & $.50$ & $2.93$ & $.098$ & $.09$ & $0.02$ & $.877$ & $.00$ \\
 Concept-tool correlation& $33.39$ & $ < .001 $ & $.53$ & $1.98$ & $.171$ & $.06$ & $1.20$ & $.284$ & $.04$ \\
 Learning motivation & $36.25$ & $ < .001 $ & $.55$ & $0.49$ & $.490$ & $.02$ & $0.18$ & $.678$ & $.01$ \\
 Adjacent concept exploration & $41.20$ & $ < .001 $ & $.58$ & $10.30$ & $.003$ & $.26$ & $0.16$ & $.691$ & $.01$ \\
 \bottomrule
 \end{tabular}
 }
\end{table}

\begin{table}[!ht]
 \centering
 \caption{Two-way ANOVA results for user interface experience measures.}
 \label{tab:anova_interfaceexperience}
 \scalebox{0.95} {
 \begin{tabular}{p{0.27\textwidth} ccc ccc ccc}
 \toprule
 & \multicolumn{3}{c}{\textbf{Interface Effect}} & \multicolumn{3}{c}{\textbf{Task Effect}} & \multicolumn{3}{c}{\textbf{Interaction Effect}} \\
 \cmidrule(lr){2-4} \cmidrule(lr){5-7} \cmidrule(lr){8-10}
 & $F_{(1, 30)}$ & $p$ & $\eta^{2}_{p}$& $F_{(1, 30)}$ & $p$ & $\eta^{2}_{p}$ & $F_{(1, 30)}$ & $p$ & $\eta^{2}_{p}$ \\
 \midrule
 Tool support vs. distraction & $44.44$ & $ < .001 $ & $.60$ & $0.44$ & $.510$ & $.01$ & $0.00$ & $1.000$ & $.00$ \\
 Prior knowledge requirement & $32.36$ & $ < .001 $ & $.52$ & $11.65$ & $.002$ & $.28$ & $2.14$ & $.155$ & $.07$ \\
 External help requirement& $26.75$ & $ < .001 $ & $.47$ & $15.35$ & $.001$ & $.34$ & $4.15$ & $.051$ & $.12$ \\
 Adjacent tool exploration & $59.25$ & $ < .001 $ & $.66$ & $3.25$ & $.082$ & $.10$ & $0.67$ & $.420$ & $.02$ \\
 Interface vs. user adaptation& $104.14$ & $ < .001 $ & $.78$ & $0.14$ & $.708$ & $.00$ & $1.29$ & $.266$ & $.04$ \\
 \bottomrule
 \end{tabular}
 }
\end{table}

\clearpage
\section{Scaffolded Interface Variations} \label{sec:interface_variations}
\subsection{UV Unwrapping}
Figure~\ref{fig:task1_interface_vars} shows the scaffolded interface variations for UV unwrapping. 
For \textit{organic} models like characters and creatures, the workflow focuses on creating smooth, low-distortion texture maps. The interface helps users define seams to flatten curved surfaces, apply algorithms that preserve surface flow, and refine UV layouts to keep textures accurate during animation.
For \textit{hard-surface} models, the workflow is designed to handle their unique geometry and texturing needs. The interface supports multiple UV sets for different textures, uses precise projection and alignment tools to create clean UV islands, and includes tools for accurate decal placement.

\subsection{Building Walk Cycle}
\begin{figure*}[!ht]
    \centering
    \begin{subfigure}[b]{0.48\textwidth}
        \includegraphics[width=\textwidth]{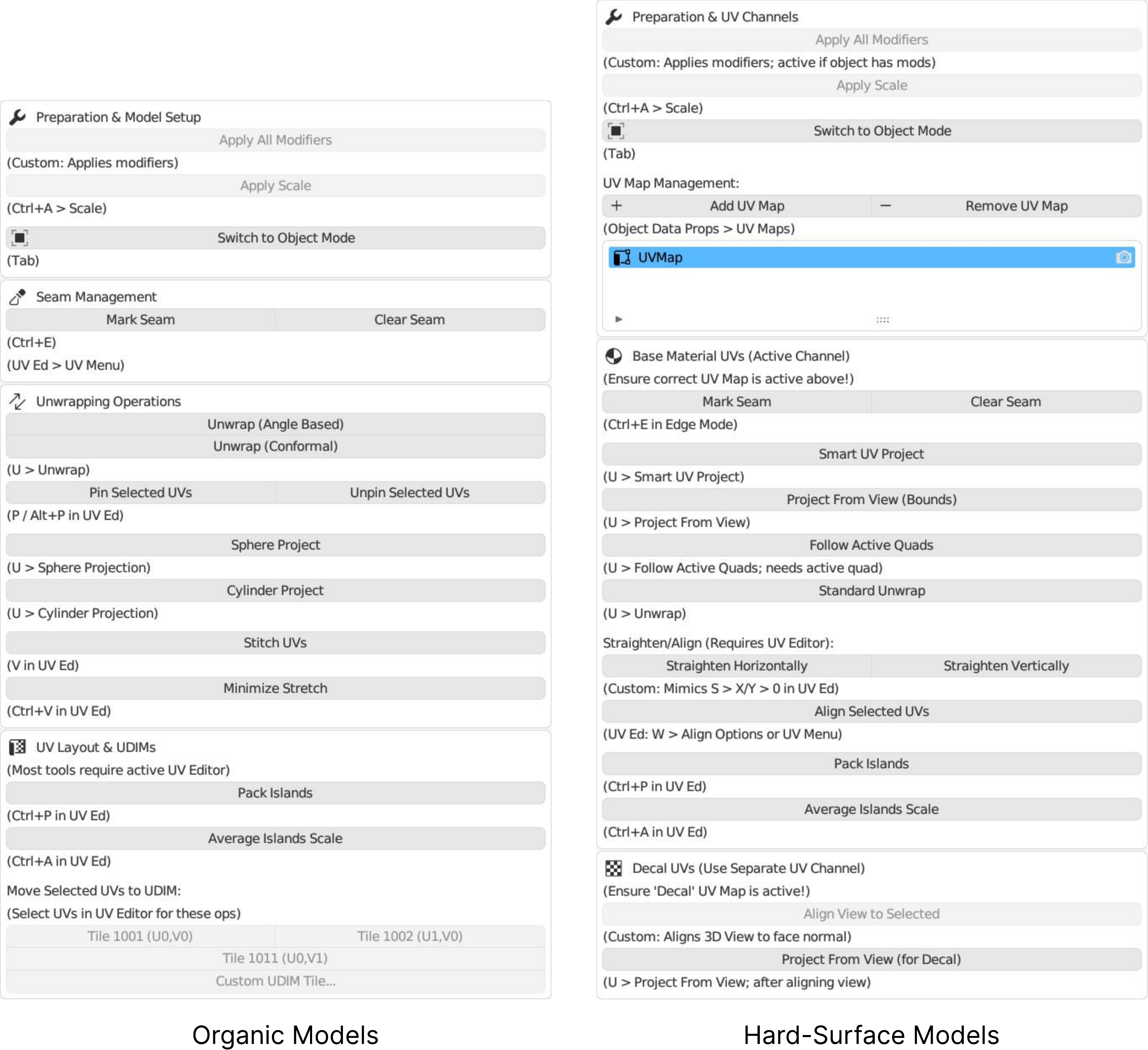}
        \caption{UV unwrapping.}
        \label{fig:task1_interface_vars}
    \end{subfigure}
    \hfill
    \begin{subfigure}[b]{0.48\textwidth}
        \includegraphics[width=\textwidth]{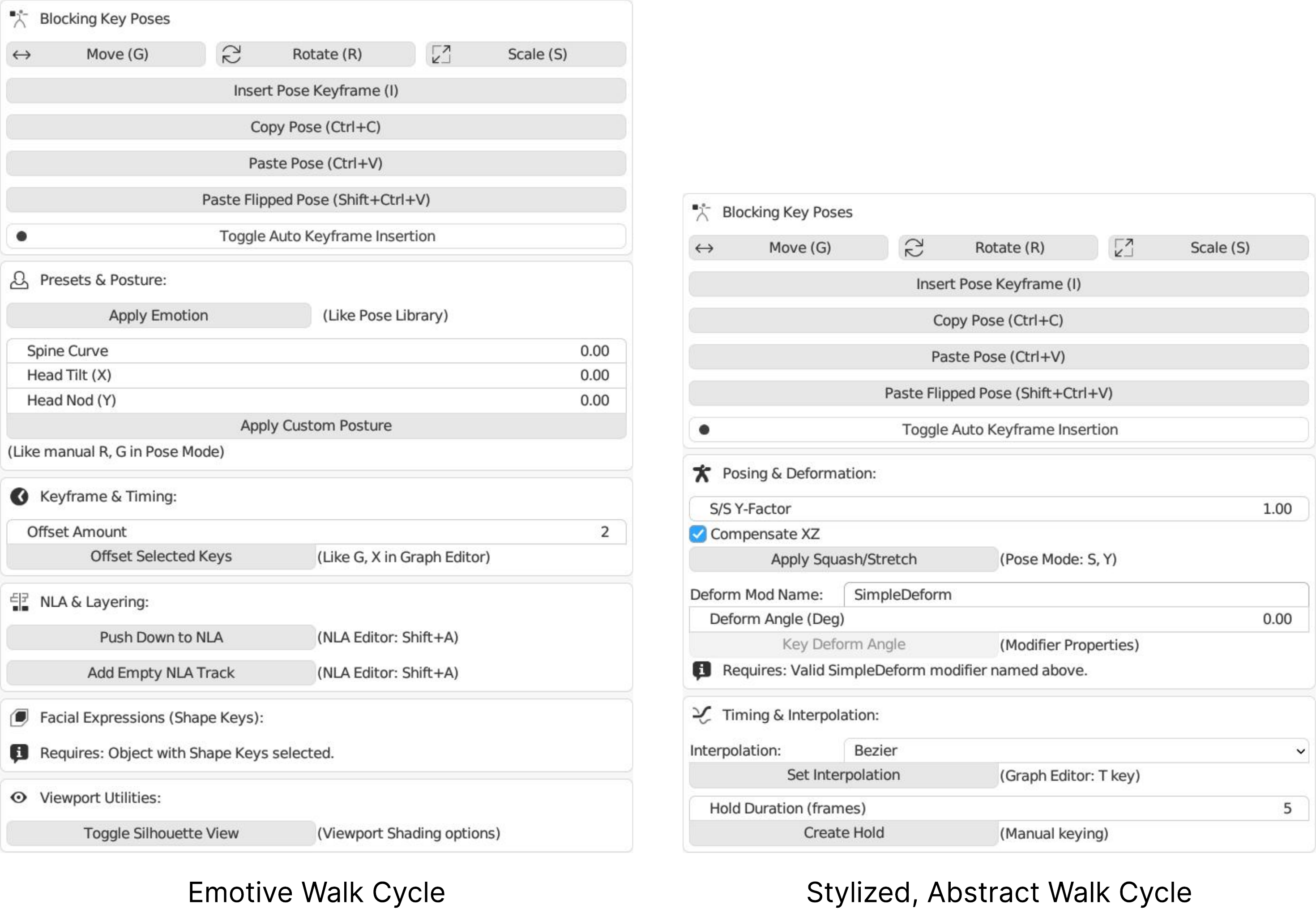}
        \caption{Building walk cycle.}
        \label{fig:task2_interface_vars}
    \end{subfigure}
    \caption{Scaffolded UI variations in Blender implemented using \papername{}.}
    \Description{This figure shows the UI variations of the UV unwrapping and walk animation tasks.}
    \label{fig:interface_vars}
\end{figure*}

Figure~\ref{fig:task2_interface_vars} presents the scaffolded interface variations for building a walk cycle. 
The \textit{emotive walk cycle} workflow adds emotion and personality to walk animations. The interface guides users in analyzing reference material, adjusting key poses, especially spine and head, and refining timing and curves, with layered secondary motion to express the desired mood or personality.
The \textit{stylized, abstract walk cycle} workflow focuses on creating animations that move away from realism for artistic effect. The interface supports exaggerating form and motion, adjusting timing and interpolation for unique rhythms, and creating bold, non-realistic effects to match artistic or conceptual goals.

\section{Scaffolded Interface in Other Software} \label{sec:interface_maya}
\begin{figure}
    \centering
    \begin{subfigure}[b]{0.3\textwidth}
        \includegraphics[width=\textwidth]{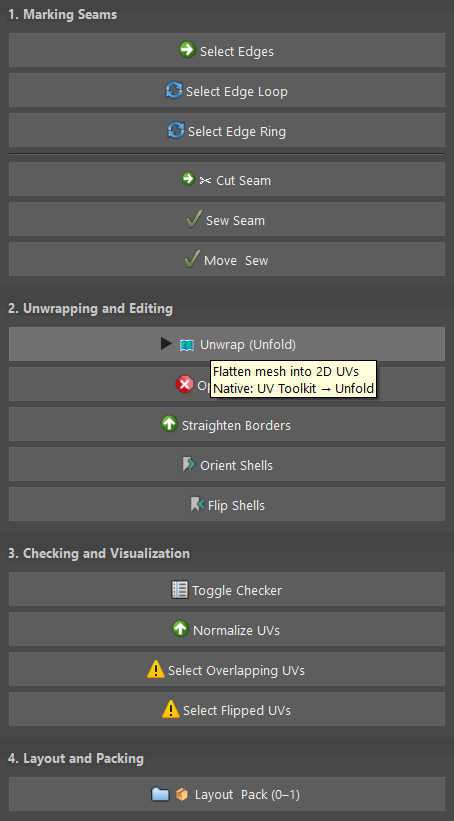}
        \caption{UV unwrapping.}
        \label{fig:task1_interface_maya}
    \end{subfigure}
    \hspace{0.5cm}
    \begin{subfigure}[b]{0.3\textwidth}
        \includegraphics[width=\textwidth]{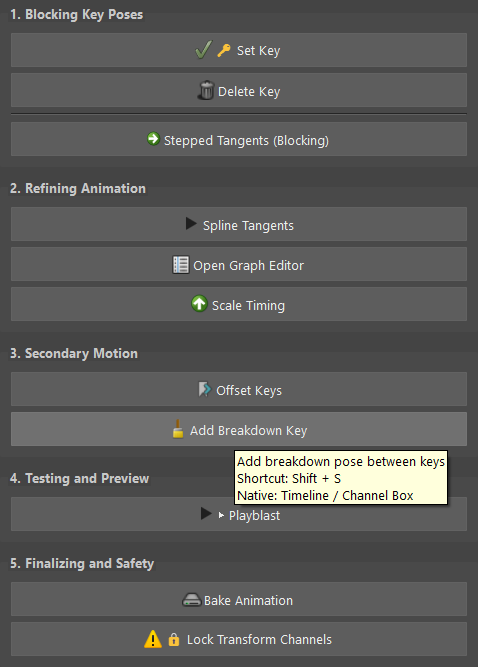}
        \caption{Building walk cycle.}
        \label{fig:task2_interface_maya}
    \end{subfigure}
    \caption{Scaffolded UIs in Maya implemented using \papername{}.}
    \Description{This figure shows the scaffolded UIs for the UV unwrapping and walk animation tasks in Maya.}
    \label{fig:interface_maya}
\end{figure}

Figures~\ref{fig:task1_interface_maya} and~\ref{fig:task2_interface_maya} demonstrate the scaffolded interfaces we implemented in Maya to show the possibility of applying our scaffolded UI generation method in other professional software. 

\end{document}